\documentclass[10pt]{article}
\usepackage{cite}
\usepackage{float}
\usepackage{times}
\usepackage{graphicx}
\usepackage{caption}
\usepackage{amsmath}
\usepackage{dblfloatfix}
\usepackage{color} 
\usepackage{multicol}

\title{Two-dimensional Ga$_2$O$_3$ glass:\\ a large scale passivation and protection material for monolayer WS$_2$} 

\author
{Matthias Wurdack,$^{1,2\ast}$ Tinghe Yun,$^{1,3}$ Eliezer Estrecho,$^{1, 2}$ Nitu Syed,$^{4}$\\Semonti Bhattacharyya,$^{1,5}$ Maciej Pieczarka,$^{1, 2}$ Ali Zavabeti,$^{4, 6}$ Shao-Yu Chen,$^{1,5}$   \\Benedikt Haas,$^{7}$ Johannes M\"uller,$^{7}$ Qiaoliang Bao,$^{3}$ Christian Schneider,$^{8}$\\ Yuerui Lu,$^{1,9}$ Michael S. Fuhrer,$^{1,5}$ Andrew G. Truscott,$^{10}$ Torben Daeneke,$^{1,4}$ \\and Elena A. Ostrovskaya$^{1, 2\ast}$ \\
\\
\normalsize{$^{1}$ ARC Centre of Excellence in Future Low-Energy Electronics Technologies}\\
\normalsize{$^{2}$ Nonlinear Physics Centre, Research School of Physics, The Australian National University,}\\
\normalsize{Canberra, ACT 2601, Australia}\\
\normalsize{$^{3}$ Department of Materials Science and Engineering, Monash University, Clayton, Australia}\\
\normalsize{$^{4}$ Department of Chemical and Environmental Engineering, RMIT, Melbourne, Australia}\\
\normalsize{$^{5}$ School of Physics and Astronomy, Monash University, Clayton, Australia}\\
\normalsize{$^{6}$ Department of Chemical Engineering, The University of Melbourne, Parkville, VIC, 3010, Australia,}\\
\normalsize{$^{7}$ Institut fur Physik \& IRIS Adlershof, Humboldt-Universit\"at zu Berlin, D-10099 Berlin, Germany}\\
\normalsize{$^{8}$ Institut of Physics, Carl von Ossietzky University of Oldenburg,}\\
\normalsize{Ammerl\"ander Heerstrasse 114-118, 26126 Oldenburg, Germany}\\
\normalsize{$^{9}$ Research School of Electrical, Energy and Materials Engineering,}\\
\normalsize{College of Engineering and Computer Science,}\\
\normalsize{The Australian National University, Canberra, ACT 2601, Australia}\\
\normalsize{$^{10}$ Laser Physics Centre, Research School of Physics, The Australian National University,}\\
\normalsize{Canberra, ACT 2601, Australia}\\
\\
\normalsize{$^\ast$ Corresponding authors. E-mail: matthias.wurdack@anu.edu.au, elena.ostrovskaya@anu.edu.au}
}

\date{}

\begin{document} 

% Double-space the manuscript.

%\baselineskip24pt

% Make the title.

\maketitle 
% Place your abstract within the special {sciabstract} environment.

\begin{multicols}{2}
\textbf{Atomically thin transition metal dichalcogenide crystals (TMDCs) have extraordinary optical properties that make them attractive for future optoelectronic applications. Integration of TMDCs into practical all-dielectric heterostructures hinges on the ability to passivate and protect them against necessary fabrication steps on large scales. Despite its limited scalability, encapsulation of TMDCs in hexagonal boron nitride (hBN) currently has no viable alternative for achieving high performance of the final device. Here, we show that the novel, ultrathin $\textrm{Ga}_{2}\textrm{O}_{3}$ glass is an ideal centimeter-scale coating material that enhances optical performance of the monolayers and protects them against further material deposition. In particular, $\textrm{Ga}_{2}\textrm{O}_{3}$ capping of commercial grade WS$_2$ monolayers outperforms hBN in both scalability and optical performance at room temperature. These properties make Ga$_2$O$_3$ highly suitable for large scale passivation and protection of monolayer TMDCs in functional heterostructures.}
\newline \indent Two-dimensional (2D), atomically thin monolayers of transition metal dichalcogenide crystals (TMDCs) are highly optically active, direct bandgap semiconductors that have emerged as a promising platform for future low-energy electronics, optoelectronics, and photonics \cite{RN258, RN481, RN576, RN256}. Extensive research points to an exceptional potential of TMDC excitons (stable electron-hole pairs) \cite{RN255} for ultra-efficient energy and information technologies \cite{RN362, RN463}, sensing \cite{RN462}, and fundamental studies of collective quantum phenomena \cite{RN468,RN469}. However, integration of monolayer TMDCs into useful electrical and optical devices by direct material deposition, e.g. of high-$\kappa$ dielectric materials for top-gating \cite{RN564, RN500, RN501}, usually degrades their electronic and optical properties. High optical and electronic performance can be achieved and retained \cite{RN571,VM2019} by full encapsulation in mechanically exfoliated hexagonal boron nitride (hBN), commonly used to passivate and protect the monolayers. However, this approach is non-scalable due to the limited size and inconsistent thickness of the exfoliated hBN flakes. Significant effort has been directed towards increasing the size of monolayer TMDCs from several $\mu\mathrm{m}$ to the $\mathrm{cm}$-scale \cite{RN489, RN460}, and the capability to passivate and protect the monolayers on similar scales is equally important. Without a large scale passivation and protection technology, the realization of multilayer structures with integrated monolayer TMDCs remains challenging and inherently non-scalable.
\begin{figure*}[hb!]
\centering
\vspace*{-0.7 cm}
 \includegraphics[width=19cm]{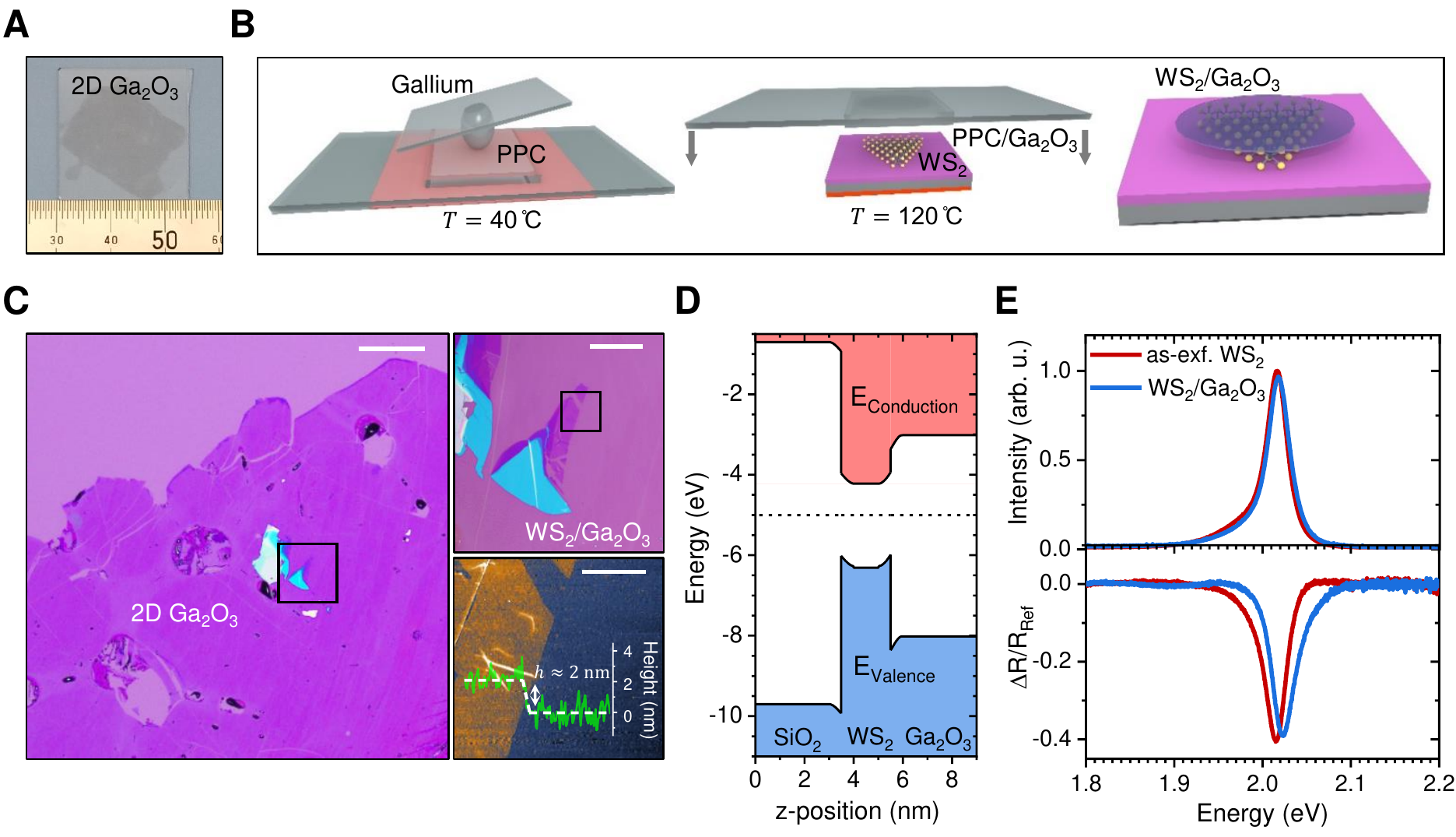}
 \caption*{{\bf Fig. 1: Large scale passivation of monolayer WS$_2$ with 2D Ga$_2$O$_3$:} (A) Camera image of a $\mathrm{cm}$-sized 2D Ga$_2$O$_3$ sheet on glass. The values on the ruler are in the $\mathrm{mm}$ scale; (B) Schematics of the PPC-assisted deterministic transfer technique for TMDC/Ga$_2$O$_3$ heterostructures (from left to right); (C) Microscope and AFM images of a WS$_2$/Ga$_2$O$_3$ heterostructure (scale bar sizes: $200~\mathrm{\mu m}$, $40~\mathrm{\mu m}$, and $10~\mathrm{\mu m}$). The extension of the dashed line at $h=0~\mathrm{nm}$ marks the position of the line profile; (D) Schematic band diagram of WS$_2$/Ga$_2$O$_3$ heterostructure, (E) PL and reflectivity spectra of an as-exfoliated monolayer WS$_2$ and of a WS$_2$/Ga$_2$O$_3$ heterostructure under ambient conditions.}
 \label{fig:figure1}
\end{figure*}
\newline \indent A practical protection and passivation material should: a) have a uniform, $\mathrm{nm}$-scale thickness on wafer scale, b) have no negative effects on the optical and electrical properties of monolayer TMDCs, and c) protect against further material deposition to enable the integration into multilayer heterostructures. Here, we introduce an isotropic, 2D $\textrm{Ga}_{2}\textrm{O}_{3}$ glass \cite{RN453} as a novel material for low-cost passivation and protection of atomically-thin semiconductors. The wide-bandgap 2D Ga$_2$O$_3$ glass can be synthesized under ambient conditions, with a highly reproducible thickness of less than $3~\mathrm{nm}$ on $\mathrm{cm}$-scale. By capping the notoriously fragile $\textrm{WS}_{2}$ with 2D Ga$_2$O$_3$, we demonstrate its excellent passivating properties. Our measurements at cryogenic temperatures indicate that the 2D Ga$_2$O$_3$ passivates the commercial grade $\textrm{WS}_{2}$ monolayer by filling in sulphide vacancies. The passivated WS$_2$ exhibits enhanced exciton photoluminscence (PL) and suppressed exciton annihilation processes, similarly to the effects of superacid treatment \cite{RN479} and full hBN encapsulation \cite{RN473}. Finally, we show that $\textrm{Ga}_{2}\textrm{O}_{3}$ protects $\textrm{WS}_{2}$ against further deposition of Al$_2$O$_3$, a high-$\kappa$ dielectric material. Comparison of the exciton PL of commercial grade monolayer WS$_2$ capped by either Ga$_2$O$_3$ or hBN shows that the 2D $\textrm{Ga}_{2}\textrm{O}_{3}$ glass outperforms hBN as a protective material at room temperature.
\newline \indent Amorphous Ga$_2$O$_3$ is an electrically insulating, isotropic glass, fully transparent in the visible range \cite{RN495}, and is therefore well suited for protection of optically active materials. The recently discovered liquid metal printing method \cite{RN453} enables the low-cost synthesis of $\mathrm{cm}$-sized 2D sheets of Ga$_2$O$_3$ (Fig. 1A) with a reproducible thickness of $\sim 3 ~\mathrm{nm}$ in ambient conditions [see Supplementary Materials S1]. X-ray photoelectron spectroscopy (XPS) on the 2D Ga$_2$O$_3$ confirms the purity of its stoichiometric composition with 60\% oxygen and 40\% gallium [see Supplementary Materials S2]. A scanning electron microscope with a transmission diffraction stage was used to measure the crystal structure of synthesized 2D Ga$_2$O$_3$, revealing that it is an entirely amorphous, isotropic glass [see Supplementary Materials S3]. Electron energy loss spectroscopy (EELS) measurements unveiled a bandgap of around $5~\mathrm{eV}$  [see Supplementary Materials S4], which is significantly wider than the bandgap of amorphous bulk Ga$_2$O$_3$ \cite{RN495}. 
\begin{figure*}[hb!]
\centering
\vspace*{-0.7 cm}
 \includegraphics[width=19cm]{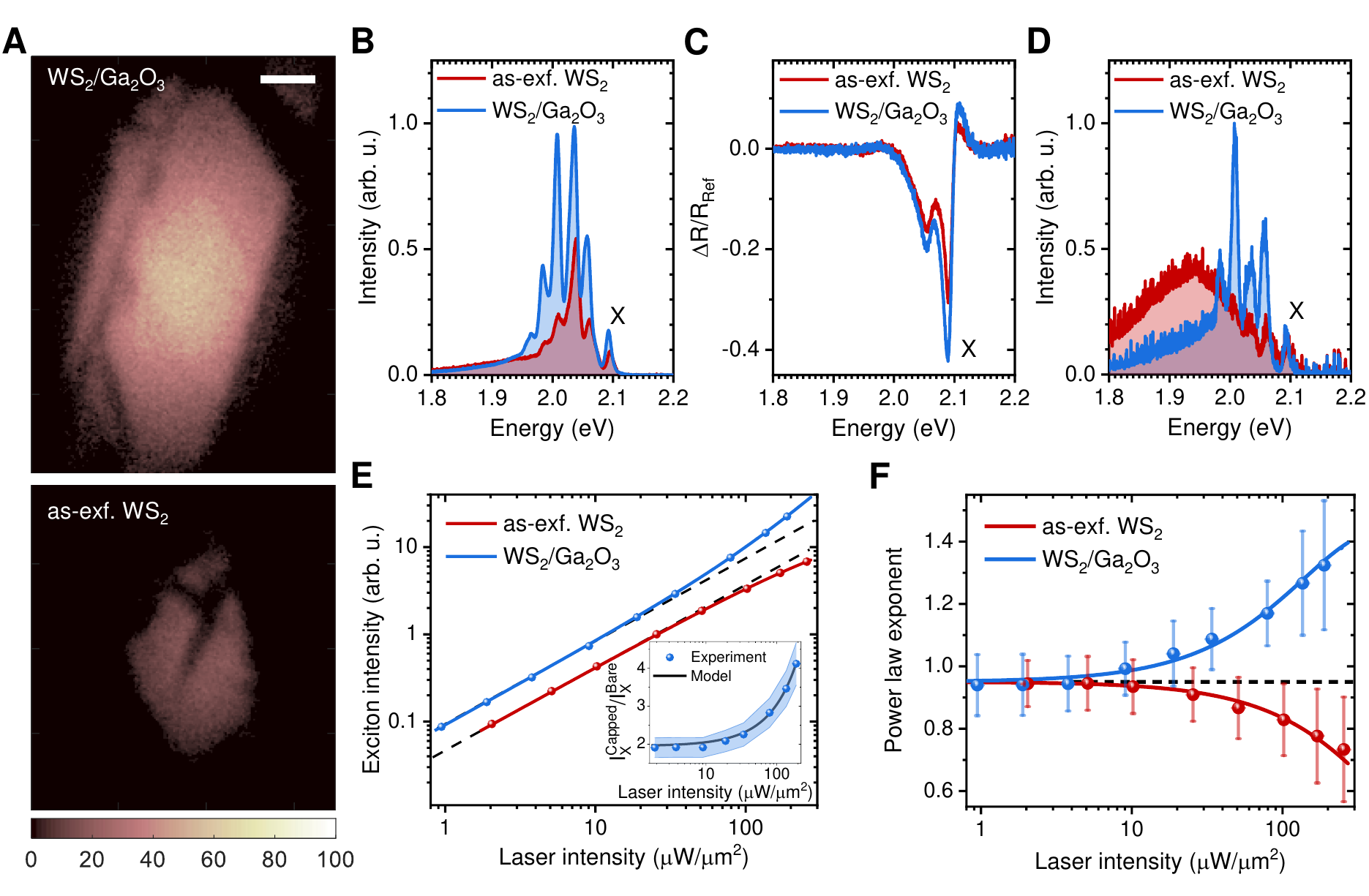}
 \caption*{{\bf Fig. 2: PL studies on a WS$_2$/Ga$_2$O$_3$ heterostructure at $\boldsymbol{T=4.3\ \mathrm{K}}$:} (A) PL images of a WS$_2$/Ga$_2$O$_3$ heterostructure and of an as-exoliated monolayer WS$_2$ under large Gaussian laser spot excitation ($\sim25~\mu m$), the scale bar size is $5\ \mathrm{\mu m}$; (B) corresponding PL spectra with $I_L\approx 34~\mathrm{\mu W/ \mu m^2}$, (C) reflectivity spectra, and (D) PL spectra with $I_L\approx 1.7~\mathrm{\mu W/ \mu m^2}$; (E) Laser intensity dependent exciton PL intensities (dots) fitted with the model from the Supplementary Materials S12 (solid line); (inset) Exciton PL intensity ratios in the bare and the capped monolayers; (F) Extracted power law-exponent $k$ (slope) for the exciton PL intensities. The dashed lines correspond to a power law exponent of $k= 0.95$.}
 \label{fig:figure2}
\end{figure*}
\newline \indent We have developed two high-yield techniques for capping monolayer TMDCs with 2D Ga$_2$O$_3$. The first technique is the direct synthesis of Ga$_2$O$_3$ on top of the monolayers, capable of covering large-area TMDCs grown by chemical vapor deposition (CVD) [see Supplementary Materials S5 and S6]. The second technique (Fig. 1B) is the deterministic transfer of 2D Ga$_2$O$_3$, synthesized on spin-coated poly-propylene carbonate (PPC), onto target areas, such as $\mu$m-sized mechanically exfoliated monolayers [Supplementary Material S7]. These methods were used to create the WS$_2$/Ga$_2$O$_3$ heterostructures with both CVD-grown  and exfoliated monolayers. Monolayer WS$_2$ features the largest band gap in the TMDC family \cite{RN256}, and therefore enables us to test whether 2D Ga$_2$O$_3$ glass sufficiently confines the excited charge carriers in the TMDCs. In what follows, we focus on monolayer WS$_2$ mechanically exfoliated from single-crystalline commercial grade bulk crystals, which have superior optical quality compared to the large-scale CVD-grown monolayers used in this work.
\newline \indent Fig. 1C shows a clean $\mathrm{mm}$-scale sheet of Ga$_2$O$_3$ deterministically transferred on top of an exfoliated monolayer WS$_2$ [see Supplementary Material S8 for the results on CVD-grown WS$_2$]. The AFM image of the sample surface shows a homogeneous monolayer coverage, with the height profile demonstrating a step size of $\sim2~\mathrm{nm}$ from the 2D Ga$_2$O$_3$ sheet to the heterostructure. The height difference to the thickness of a bare monolayer WS$_2$ ($\sim 1 \ \mathrm{nm}$ \cite{RN256}) can be attributed to a pronounced van der Waals spacing between the monolayer and the Ga$_2$O$_3$ \cite{RN499}. The XPS and EELS measurements shown in Supplementary Materials S1 and S4, together with the known material properties of monolayer WS$_2$ \cite{RN575, RN563}, amorphous SiO$_2$ \cite{RN498}, and amorphous Ga$_2$O$_3$ \cite{RN495}, allow us to construct a band diagram of the WS$_2$/Ga$_2$O$_3$ heterostructure (see Fig. 1D), while the exact position of the WS$_2$ band edge is still under debate  \cite{RN575, RN255, RN563}. The identified type I band alignment enables the Ga$_2$O$_3$ to confine the free carriers in the conduction and valence bands of WS$_2$, to form excitons.
\newline \indent To test the effect of the Ga$_2$O$_3$ capping on the exciton properties of monolayer WS$_2$, we performed PL and reflectivity studies. The system was excited by a ND:YAG continuous wave (cw) laser source with a wavelength of $\lambda = 532~\mathrm{nm}$ ($E  \approx2.33~ \mathrm{eV}$) which is energetically above the band gap of monolayer WS$_2$, but below the band gap of 2D Ga$_2$O$_3$. The reflectivity measurements were performed with a tungsten halogen white light source. The PL and reflectivity spectra of an as-exfoliated monolayer WS$_2$ and of the WS$_2$/Ga$_2$O$_3$ heterostructure under ambient conditions feature the exciton PL and absorption at $E\approx 2.01\ \mathrm{eV}$ (Fig. 1E). The excitons in WS$_2$/Ga$_2$O$_3$ have slightly higher energies compared to the as-exfoliated WS$_2$, most likely due to dielectric screening effects in Ga$_2$O$_3$. Nevertheless, the amplitude and the linewidth of the PL and reflectivity spectra are not affected by Ga$_2$O$_3$ capping, indicating that the WS$_2$ excitons are not quenched. The PL and Raman measurements on the CVD-grown monolayers confirm these results, and show that the optical phonon modes of monolayer WS$_2$ remain intact [see Supplementary Materials S8]. By contrast, exciton PL of exfoliated WS$_2$ monolayers passivated with the commercial grade hBN is significantly quenched [see Supplementary Materials S9].
\newline \indent
The effect of the Ga$_2$O$_3$ capping on the WS$_2$ exciton PL and absorption was further tested at $T = 4.3 ~\mathrm{K}$, when exciton-phonon interactions and thermal effects are suppressed. We excited the as-exfoliated WS$_2$ and the heterostructure with a large Gaussian laser spot of the diameter \linebreak$\sim25~\mathrm{\mu m}$, which effectively eliminated artefacts due to the sample inhomogeneities by averaging the PL over a large sample area. Fig. 2A contains the corresponding PL intensity maps, showing that both the as-exfoliated WS$_2$ and the heterostructure have a homogeneous PL texture. The PL spectra of the monolayers (Fig. 2B) are composed of multiple distinct PL peaks, including the exciton ($E_X \approx 2.09 \mathrm{eV}$), the trion ($E_T \approx 2.06 \mathrm{eV}$) and additional low energy peaks associated with many-body complexes \cite{RN256, RN502,RN503}. The reflectivity spectra (Fig. 2C) of the exfoliated monolayers show strong absorption features at the exciton and trion energies, which indicates that both the uncapped and the capped samples are doped \cite{RN571}. Since the ratios between the absorption dips for both samples are the same, this effect is not related to the Ga$_2$O$_3$ capping and can be traced to the large density of sulphide vacancies in the commercial grade monolayer WS$_2$ resulting in high intrinsic n-type doping \cite{RN574}.
\newline \indent The PL intensity of the heterostructure is strongly enhanced at cryogenic temperatures (see Fig. 2A,B and Supplementary Materials S8, S10). In particular, at low excitation intensities, the broad bound exciton peak, which dominates the PL spectrum of the as-exfoliated sample around $146~\mathrm{meV}$ below the free exciton energy \cite{RN577}, is strongly quenched while the PL of the excitons, trions and low-energy states are enhanced (Fig. 2D). This indicates that Ga$_2$O$_3$ capping suppresses formation of bound excitons at sulphide vacancies \cite{RN505}, enhancing the formation of many-body states and the total exciton quantum yield. Indeed, cathodoluminescene (CL) measurements at $T=70~ \mathrm{K}$ presented in Supplementary Materials S11, reveal the presence of deep donor-type oxide vacancies in Ga$_2$O$_3$ \cite{RN506} with a higher concentration of vacancies in the WS$_2$/Ga$_2$O$_3$ heterostructure. This result suggests that the 2D Ga$_2$O$_3$ fills the sulphide vacancies in the WS$_2$ with oxides, which creates donor-type oxide vacancies in the Ga$_2$O$_3$. This mechanism can explain the suppression of bound exciton formation in the WS$_2$/Ga$_2$O$_3$ without affecting its doping level.
\newline \indent To understand the origins of the PL enhancement, we measured the exciton PL of both the as-exfoliated WS$_2$ and the WS$_2$/Ga$_2$O$_3$ heterostructure for a range of laser intensities $I_L$ spanning two orders of magnitude (Fig. 2E). The corresponding PL spectra are shown in Supplementary Materials S10. At laser intensities below $\sim10~\mu$W/$\mu$m$^{2}$, the exciton intensity of the uncapped monolayers follows the power law, $I_{X}  \propto I_L^{k}$, shown by the dashed lines in Fig. 2E.  Without losses, and assuming the direct photon-exciton transition, the dependence is linear: $k=1$ \cite{RN554}. In our samples, at low laser intensities, $k \approx 0.95$, which indicates that the exciton formation experiences intensity-dependent losses, e.g., free-to-bound exciton transitions or formation of many-body exciton complexes (Fig. 2D). However, at laser intensities above $\sim10~\mu$W/$\mu$m$^{2}$, the exponent of the power law $k \propto \log I_X/\log I_L$ decreases for the as-exfoliated WS$_2$ (see Fig. 2F), in line with previous observations \cite{RN473}. This behaviour is typically caused by annihilation processes between multiple excitons or between excitons and defects \cite{RN555}, e.g. Auger recombination \cite{RN490}. 
\newline \indent In contrast, for the heterostructure the power law exponent remains well above $k\approx 0.95$. This indicates that the exciton annihilation in monolayer WS$_2$ is suppressed by Ga$_2$O$_3$ capping, similarly to the effect of full hBN encapsulation \cite{RN473,RN490}. However, the mere suppression of annihilation would result in a linear power law with a constant $k\approx 1$ \cite{RN473}, while in WS$_2$/Ga$_2$O$_3$ the exciton PL transitions from a linear to a nonlinear process with increasing laser intensities (Fig. 3F). This behaviour can be explained with a two-step exciton generation process via the deep oxide vacancies in the 2D Ga$_2$O$_3$ [see Supplementary Materials S12], in addition to the direct electron-hole excitation. The indirect process involves electron tunnelling between the donor-type oxide vacancies and the WS$_2$ valence  band as an intermediate step in the WS$_2$ exciton formation. This results in accumulation of exciton densities and growing enhancement of the exciton PL intensity. \begin{figure}[H]
\vspace*{-0.2 cm}
 \includegraphics[width=9.417cm]{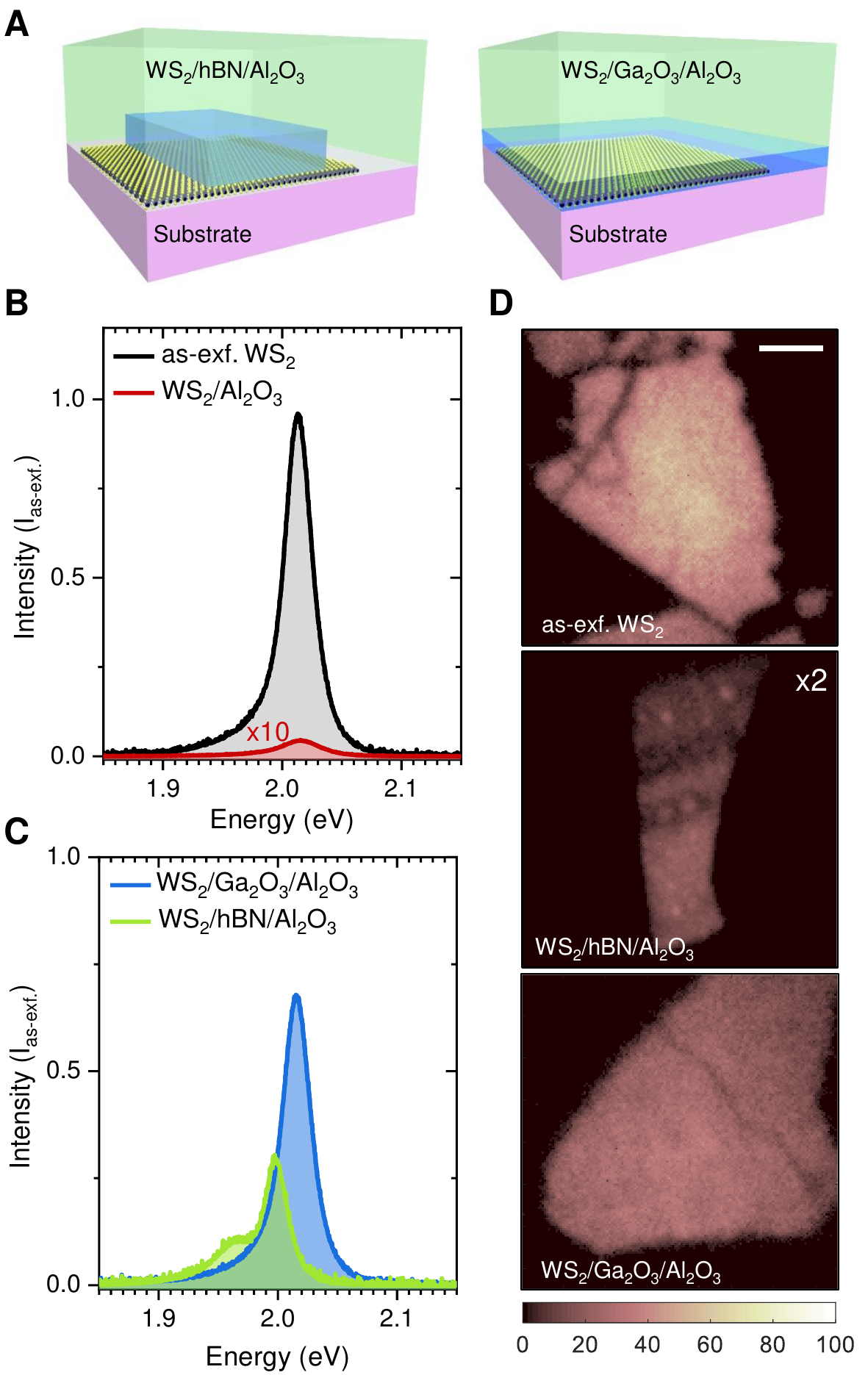}
 \caption*{{\bf Fig. 3: Integration of monolayer WS$_2$ in a high-$\kappa$ dielectric environment:} (A) Schematics of WS$_2$/hBN/Al$_2$O$_3$ and WS$_2$/Ga$_2$O$_3$/Al$_2$O$_3$ heterostructures on SiO$_2$ substrates; (B) PL spectra of monolayer WS$_2$ on SiO$_2$ before (as-exfoliated) and after Al$_2$O$_3$ deposition by EBE;(C) PL spectra of WS$_2$/hBN/Al$_2$O$_3$ and WS$_2$/Ga$_2$O$_3$/Al$_2$O$_3$ heterostructures; (D) PL intensity maps of as-exfoliated WS$_2$, and WS$_2$/hBN/Al$_2$O$_3$ and WS$_2$/Ga$_2$O$_3$/Al$_2$O$_3$ heterostructures under Gaussian spot excitation ($\sim 25~\mu m$). PL intensity of WS$_2$/hBN/Al$_2$O$_3$ is multiplied by a factor of 2. The scale bar size is $5~\mathrm{\mu m}$.}
 \label{fig:figure3}
\end{figure} 
\newpage By fitting the exciton intensities and the extracted power law exponents of both as-exfoliated WS$_2$ and the heterostructure (see Fig. S2E-F) with the theoretical model [see Supplementary S12], we deduce that the exciton generation rate in the heterostructure is $1.9\pm0.3$  times higher, and that the exciton saturation density dictated by the annihilation processes is $28\pm3$ times higher compared to the as-exfoliated WS$_2$. These accumulating effects lead to the observed power-dependent enhancement of the exciton PL intensity in WS$_2$/Ga$_2$O$_3$, with an enhancement factor of $4.1\pm0.6$ within our power range (Fig. 2E, inset). Measurements on CVD-grown monolayer WS$_2$ show the same PL intensity enhancement at cryogenic temperatures [Supplementary Materials S8].
\newline \indent To test the protective capacity of 2D Ga$_2$O$_3$ for device integration, we deposited Al$_2$O$_3$, a high-$\kappa$ dielectric material, onto monolayer WS$_2$  capped by either Ga$_2$O$_3$ or exfoliated commercial grade hBN (Fig. 3A). For deposition, we used electron beam evaporation (EBE) with a relatively high electron beam energy to stress-test the protection. As seen in Fig. 3B, direct EBE deposition of Al$_2$O$_3$ on top of bare WS$_2$ strongly quenches the exciton PL \cite{RN564}. The PL spectra (Fig. 3C) of the hBN and Ga$_2$O$_3$ capped WS$_2$ after EBE of Al$_2$O$_3$ show that both methods protect well against further material deposition, with approximately two orders of magnitude enhancement of exciton PL intensities compared to WS$_2$/Al$_2$O$_3$. 
\newline \indent However, the PL spectrum of the hBN-capped WS$_2$ (Fig. 3C) shows a pronounced shoulder at the trion energy ($E_T\approx 1.96~\mathrm{eV}$) which indicates that it is strongly doped after Al$_2$O$_3$ deposition [see Supplementary Materials S9]. In addition, the hBN flake covers the monolayer only partially, resulting in a relatively small protected area (Fig. 3D). In contrast, the PL spectrum of the WS$_2$/Ga$_2$O$_3$/Al$_2$O$_3$ heterostructure shows strong neutral exciton PL with around $70\%$ of the exciton intensity of an as-exfoliated WS$_2$ monolayer. The PL texture of the WS$_2$/Ga$_2$O$_3$/Al$_2$O$_3$ is homogeneous (Fig. 3D), which highlights the homogeneity of the coverage. 
The PL measurements on Ga$_2$O$_3$-capped CVD-grown monolayer WS$_2$ support these observations [see Supplementary Materials S8].
\newline \indent To summarise, the novel 2D Ga$_2$O$_3$ glass shows great potential as a low-cost and practical wide-bandgap, isotropic material for scalable passivation of monolayer WS$_2$. Capping the monolayers with Ga$_2$O$_3$, either by direct printing or by deterministic transfer, fully preserves their exciton properties in ambient conditions. At cryogenic temperatures, the Ga$_2$O$_3$ passivation significantly enhances the optical performance of WS$_2$ by suppressing bound exciton formation at the sulphide vacancies, promoting nonlinear exciton generation, and suppressing the exciton annihilation processes. These findings provide a pathway towards high-performance surface-passivated TMDC/Ga$_2$O$_3$ heterostructures on a $\mathrm{cm}$-scale, e.g., by combining the large-scale mechanical exfoliation \cite{RN460} with the Ga$_2$O$_3$ capping. Finally, our finding that 2D Ga$_2$O$_3$ glass outperforms hBN for protecting commercial grade monolayer WS$_2$ against high-$\kappa$ dielectric material deposition breaks new ground for the integration of monolayer TMDCs into large-area 2D devices.

\end{multicols}

\newpage

% Double-space the manuscript.

%\baselineskip24pt

% Make the title.

% Place your abstract within the special {sciabstract} environment.

%\section*{Acknowledgments}

%Here you should list the contents of your Supplementary Materials -- below is an example. 
%You should include a list of Supplementary figures, Tables, and any references that appear only in the SM. 
%Note that the reference numbering continues from the main text to the SM.
% In the example below, Refs. 4-10 were cited only in the SM.     
\section*{Supplementary Materials}
\subsection*{Material properties of 2D Ga$_2$O$_3$.}
\paragraph*{S1: Atomic force microscopy of 2D Ga$_2$O$_3$ and WS$_2$/Ga$_2$O$_3$ heterostructures.}
\begin{figure}[H]
\centering
 \includegraphics[width=\linewidth]{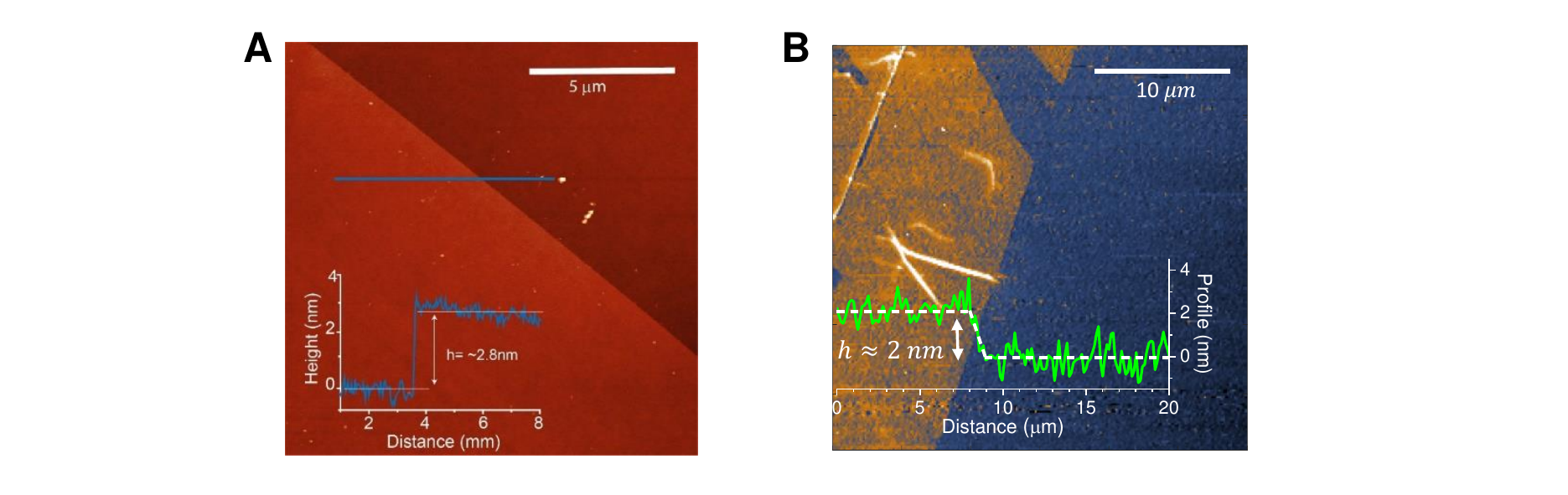}
 \caption*{{\bf Fig. S1:} Atomic force microscope (AFM) image of (A) 2D Ga$_2$O$_3$ synthesized on SiO$_2$ and (B) WS$_2$/Ga$_2$O$_3$ heterostructure on SiO2$_2$ substrate; the extension of the $h=0~\mathrm{nm}$ marks the position of the line profile.}
 \label{fig:S1}
\end{figure}
The surface morphology and thicknesses of the 2D Ga$_2$O$_3$ sheet and of WS$_2$/Ga$_2$O$_3$ were explored by atomic force microscopy (AFM). The AFM measurements were collected using a Bruker Dimension Icon AFM with Scanasyst-air AFM tips. Gwyddion 2.36 software was employed for the AFM image processing and analysis. As seen in ({\color{blue}Fig. S1A}), the typical step height from the substrate to the Ga$_2$O$_3$ is $h\sim 2.8 ~ \mathrm{nm}$. The step height from the Ga$_2$O$_3$ sheet to the heterostructure is $h\sim 2 ~\mathrm{nm}$ ({\color{blue}Fig. S1B}). 

\paragraph*{S2: X-ray photoelectron spectroscopy of 2D Ga$_2$O$_3$.}
\begin{figure}[H]
\centering
 \includegraphics[width=\linewidth]{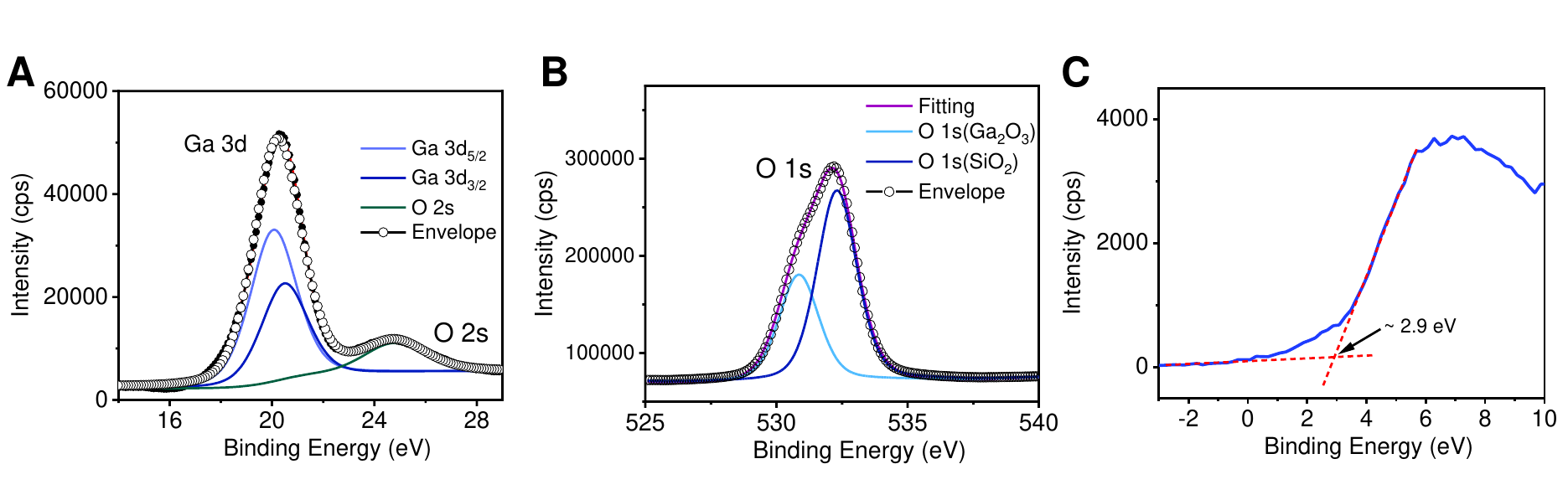}
 \caption*{{\bf Fig. S2:} X-ray photoelectron spectroscopy (XPS) of (A) the Ga 3d, and (B) the O 1s regions of 2D Ga$_2$O$_3$. (C) XPS valence band analysis of the 2D Ga$_2$O$_3$ sheet.}
 \label{fig:S2}
\end{figure}
X-ray photoelectron spectroscopy (XPS) was carried out to explore the chemical bonding states of 2D Ga$_2$O$_3$. {\color{blue}Fig. S2A-B} illustrate the spectra of Ga 3d and O 1s regions, respectively, for the Ga$_2$O$_3$. The characteristic broad peak of the Ga 3d region centred at $20.1 ~ \mathrm{eV}$, is deconvoluted into $3d_{5/2}$ and $3d_{3/2}$ components attributed to the presence of the Ga$^{3+}$ state in Ga$_2$O$_3$ \cite{RN448}. No peak is detected in the Ga 3d region associated with elemental gallium (Ga$^0$). The peak located at $\sim 25.2 ~ \mathrm{eV}$ in the Ga 3d region corresponds to the broad O 2s feature \cite{RN449}. The peak deconvolution of the O 1s presented in {\color{blue} Fig. S2B} signals the presence of two oxygen species with binding energies of $\sim 530.9 ~ \mathrm{eV}$ and $\sim 532.5 ~ \mathrm{eV}$. The peak with the higher binding energy is related to the Si-O-Si bond and the peak at $\sim 530.9~ \mathrm{eV}$ is the contribution from the Ga$_2$O$_3$ \cite{RN448}.

The valence band XPS spectrum presented in {\color{blue}Fig. S2C} reveals that the energy difference between the valence band maximum and the Fermi energy is $\sim 2.9 ~ \mathrm{eV}$. The XPS spectrum was measured using a Thermo Scientific K$_\alpha$ XPS spectrometer equipped with a monochromatic Al K$_\alpha$ source ($hv = 1486~ \mathrm{eV}$) with a spot size of approximately $\sim 400~ \mathrm{\mu m}$. All the core-level spectra (Ga 3d, O 1s, C 1s, etc.) were recorded with a pass energy of $50 ~ \mathrm{eV}$. XPS data acquisition and peak fitting analysis was done using the dedicated XPS Avantage software.

\paragraph*{S3: Crystallinity of 2D Ga$_2$O$_3$.}
\begin{figure}[H]
\centering
 \includegraphics[width=\linewidth]{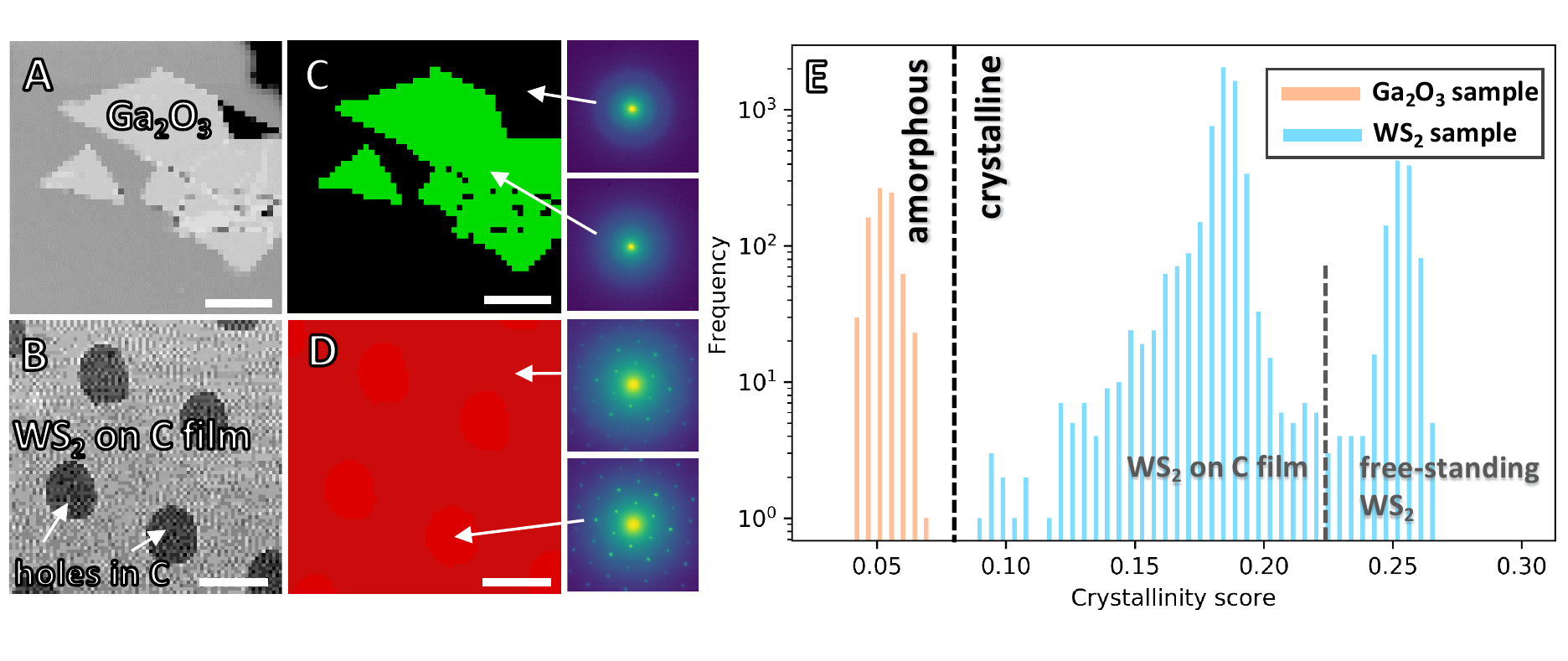}
 \caption*{{\bf Fig. S3:} Measurement of electron diffraction in 2D Ga$_2$O$_3$ and monolayer WS$_2$. (A,B) Virtual images from an electron diffraction map of a synthesized 2D Ga$_2$O$_3$ sheet as used in the main article, and of monolayer WS$_2$ on a carbon film with holes, respectively; (C, D) maps of the crystallinity of the 2D Ga$_2$O$_3$ and monolayer WS$_2$, respectively, derived from the individual diffraction patterns (right panels). The reconstruction program presents diffraction patterns from crystalline structures as red and diffraction patters from amorphous structures as green, showing that the 2D Ga$_2$O$_3$ is fully amorphous; (E) Logarithmic histogram of the algorithm output used to determine the crystallinity for 2D Ga$_2$O$_3$ and monolayer WS$_2$. Scale bars in (A) and (C) correspond to $0.5~ \mathrm{\mu m}$ and in (B) and (D) to $2~\mathrm{\mu m}$. The different shades of red in (D) correspond to values above and below the 0.225 marker in (E).}
 \label{fig:S3}
\end{figure}
To assess the crystal structure of both the synthesized 2D Ga$_2$O$_3$ and monolayer WS$_2$ used in this article, we used a modified Zeiss Gemini 500 scanning electron microscope (SEM) with a transmission diffraction stage \cite{RN450} and acquired a map of 80 by 80 electron diffraction patterns for 2D Ga$_2$ and 50 by 50 electron diffraction patterns for monolayer WS$_2$. The microscope operated at $20\  \mathrm{kV}$ for 2D Ga$_2$O$_3$ and at $15\  \mathrm{kV}$ for monolayer WS$_2$ with a beam size of roughly $4\ \mathrm{nm}$ and a step size in between patterns of around $40\ \mathrm{nm}$ for 2D Ga$_2$O$_3$ and  $100\ \mathrm{nm}$ for monolayer WS$_2$. Each pattern was recorded with an exposure time of $25\ \mathrm{ms}$.

{\color{blue}Fig. S3A-B} shows virtual images of 2D Ga$_2$O$_3$ and of a WS$_2$ monolayer. To determine whether the material is crystalline or amorphous, the diffraction patterns were analyzed for clusters of intensity (diffraction spots) with a self-made script that utilizes the Python package {\verb}trackpy} \cite{trackpy}. A histogram of the algorithm output for the two samples ({\color{blue}Fig. S3E}) shows a sharp threshold between the amorphous and crystalline phases, beyond which patterns with spots would be situated. Applying the threshold and converting the data back results in the maps in {\color{blue}Fig. S3C-D}, which show two examples for the scattering observed from the amorphous 2D Ga$_2$O$_3$ and from the single-crystalline WS$_2$. This analysis demonstrates that 2D Ga$_2$O$_3$ sheets used in our experiments are amorphous.

\begin{figure}[H]
\centering
 \includegraphics[width=\linewidth]{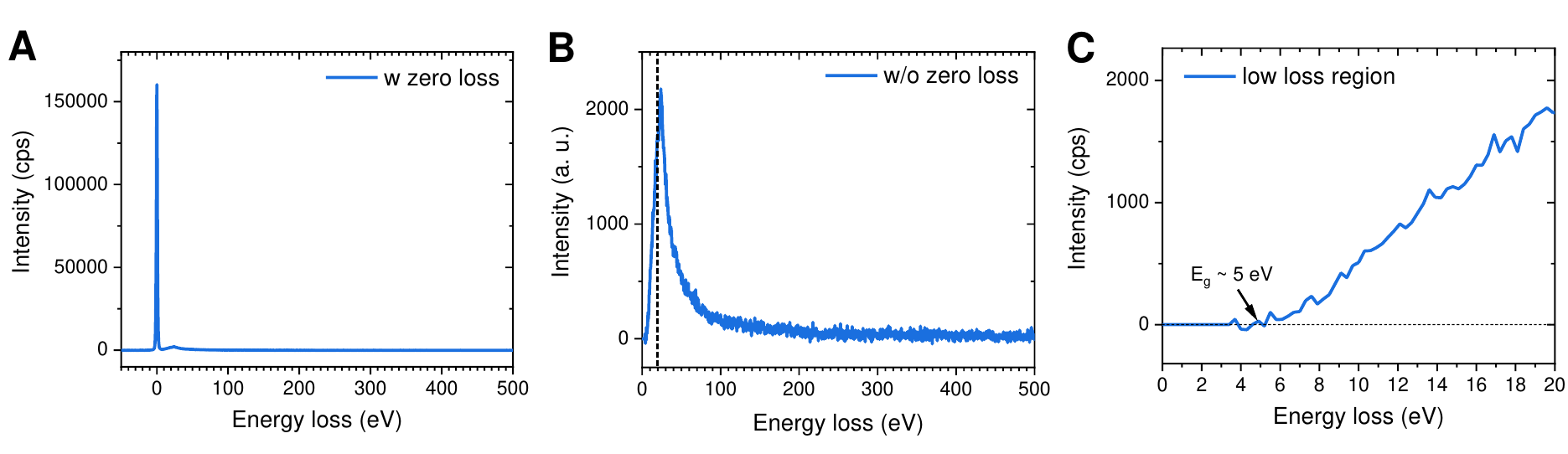}
 \caption*{{\bf Fig. S4:} Electron energy loss spectroscopy (EELS) of amorphous 2D Ga$_2$O$_3$: (A) EELS spectrum in the energy range up to $E=500\ \mathrm{eV}$, including the zero-loss peak at $E_{loss}=0~ \mathrm{eV}$, (B) EELS spectrum after subtraction of the zero-loss peak (low-loss region marked with a black dashed line), (C) EELS low-loss region from (B) unveiling an optical bandgap of $E_g \approx 5 \ \mathrm{eV}$.}
\label{fig:S4}
\end{figure}
\paragraph*{S4: Optical band gap of 2D Ga$_2$O$_3$.} To determine the bandgap of a synthesized 2D Ga$_2$O$_3$ sheet via electron energy loss spectroscopy (EELS), the Ga$_2$O$_3$ was transferred onto a transmission electron microscope (TEM) grid. For this measurement, the TEM model JEOL JEM-2100f operating at the acceleration voltage of $200 ~ \mathrm{kV}$ and equipped with the Gatan Tridium imaging filter was used. EELS measures the change in the kinetic energy of electrons after interactions with the sample, and the EELS spectrum at lower binding energies  ($E_{loss}<50~  \mathrm{eV}$) can reveal the bandgap of materials. In the low-loss region the intensity of electrons that pass through the sample without experiencing any losses dominate the spectrum ({\color{blue}Fig. S4A}). This zero-loss peak needs to be removed before the inelastic interactions between the sample and the electron beam can be analysed. Here, the zero-loss peak is centred at $0~\mathrm{eV}$ with a full-width-at-half-maximum (FWHM) of less than $1.5 ~ \mathrm{eV}$, thus featuring a reliable technique for measuring materials with a wide bandgap, including Ga$_2$O$_3$. The EELS measurement was conducted in the scanning transmission electron microscope (STEM) mode. The software Digital Micrograph v1.84 was used to analyse and remove the zero-peak loss. The resulting EELS spectrum is presented in {\color{blue}Fig. S4B} for the energy range up to $E=500~ \mathrm{eV}$. The optical bandgap of the Ga$_2$O$_3$ sheet corresponds to the onset of the EELS-spectrum in the low-loss region (here, $E<20 ~ \mathrm{eV}$)  and, according to {\color{blue}Fig. S4C}, is located at $E_g \approx 5 ~ \mathrm{eV}$. 
     
\subsection*{Fabrication of WS$_2$/Ga$_2$O$_3$ heterostructures.}
\paragraph*{S5: Fabrication and transfer of CVD-grown monolayer WS$_2$.}
\begin{figure}[H]
\centering
 \includegraphics[width=\linewidth]{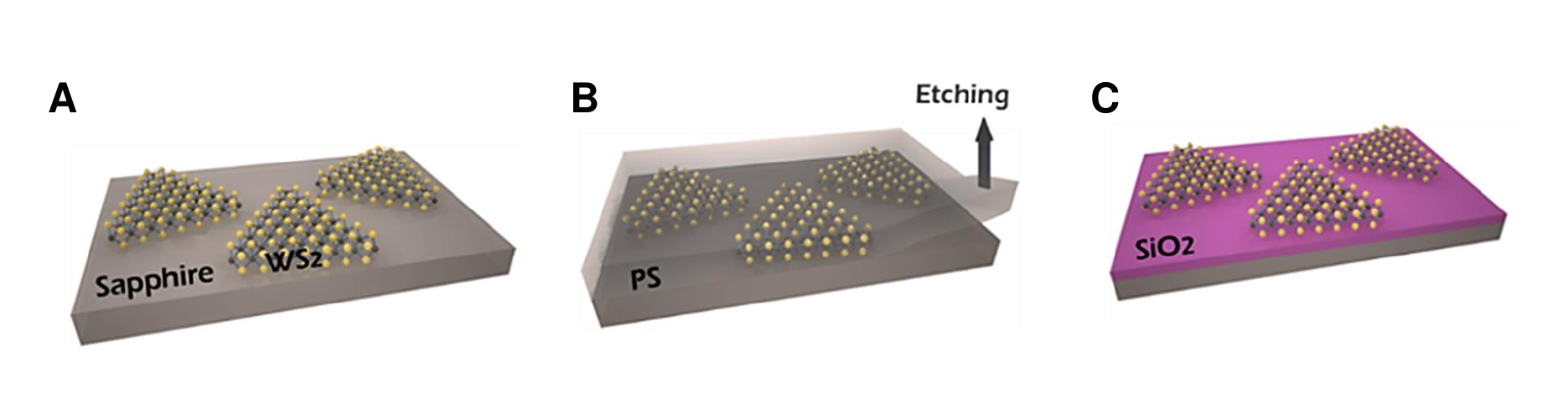}
 \caption*{{\bf Fig. S5:} Schematic illustration of the wet transfer of CVD-grown monolayers: (A) monolayers are grown on a sapphire substrate, (B) polystyrene (PS) is spin coated on top of the sapphire substrate to remove the monolayers, (C) wet transfer of the monolayers onto the SiO$_2$ substrate.}
 \label{fig:S5}
\end{figure}
Monolayer WS$_2$ crystals are grown on a sapphire [Al$_2$O$_3$ (0001)] substrate by a chemical vapor deposition (CVD) process, similar to the method reported in \cite{RN451}. Therefore, the WO$_3$ and S powders are placed in separate quartz crucibles and located at the upstream region of a tubular quartz reactor, followed by the sapphire substrate in the downstream region. Further, the WO$_3$ and S powders are heated up to $860  \ \mathrm{^{\circ}C}$ and $180 \ \mathrm{^{\circ}C}$, with the temperature rates of $28 \  \ \mathrm{^{\circ}C/min}$ and $5 \  \ \mathrm{^{\circ}C/min}$, respectively. The growth of the WS$_2$ is finished $15 \ \mathrm{min}$ after reaching these temperatures.

After the growth, monolayer WS$_2$ crystals can be transferred by means of the polymer-assisted wet transfer technique \cite{RN451}. In this work, polystyrene (Mw=192 000) (PS) in toluene solution (50 mg ml$^{-1}$) is used for the transfer onto a Si/SiO$_2$ substrate. First, PS is spin-coated on top of the CVD-grown monolayers at 3000 rpm for 60 s and then baked at $80 \ \mathrm{^{\circ}C}$ for $5\ \mathrm{min}$. After the baking process, a 2 mol L$^{-1}$ NaOH solution is used to delaminate the PS/WS$_2$ crystal film from the sapphire substrate. To remove the remaining alkali residues, the film is placed in a deionized water bath. Finally, the PS film is picked up from the bath with the Si/SiO$_2$ target substrate and dissolved in acetone, which leaves the monolayers transferred on the substrate surface. As a result, we fabricated and transferred a large area of monolayer WS$_2$ crystals on top of a Si/SiO$_2$ substrate. 

\paragraph*{S6: Direct synthesis method.}
\begin{figure}[H]
\centering
 \includegraphics[width=\linewidth]{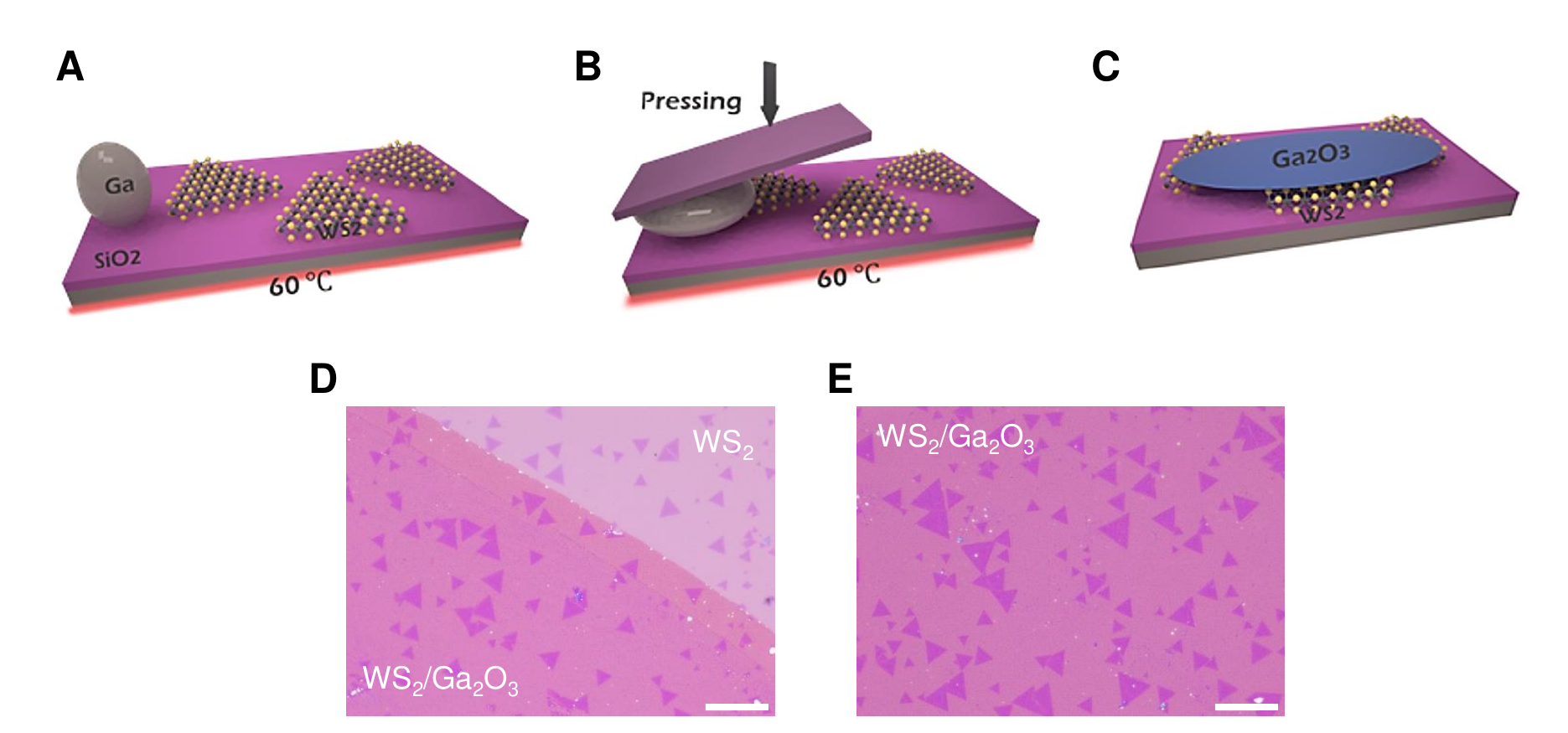}
 \caption*{{\bf Fig. S6:} Schematic illustration of the direct synthesis of 2D Ga$_2$O$_3$ on top of monolayer TMDCs: (A) a gallium droplet is placed next to TMDC area, (B) transfer of the 2D Ga$_2$O$_3$ sheet by squeezing the gallium droplet over the TMDC area, (C) 2D Ga$_2$O$_3$ sheet transferred on top of monolayer TMDCs, (D) Microscope images of the boundary and of (E) a central area of a large scale WS$_2$/Ga$_2$O$_3$ heterostructure respectively (scale bar: $40 ~\mathrm{\mu m}$), }
 \label{fig:S6}
\end{figure}
To synthesize 2D Ga$_2$O$_3$ directly on top of monolayer TMDCs, we utilized the liquid gallium printing technique for making ultrathin sheets of Ga$_2$O$_3$\cite{RN453,RN454}. Gallium has a melting point of  $29.76 \ \mathrm{^{\circ}C}$ \cite{RN455} and is therefore liquid at temperatures which are harmless for monolayer TMDCs. In air, the surface of liquid gallium transforms to a ultrathin Ga$_2$O$_3$ shell due to the self-limiting oxidation process \cite{RN453}, which is more effective at elevated temperatures. 

In this work, the samples were fabricated on a hot plate at  $60\ \mathrm{^{\circ}C}$ in air to accelerate the
oxidization process without damaging the monolayers. To start the synthesis, a small liquid gallium droplet is placed on the sample surface next to the monolayer WS$_2$ area ({\color{blue}Fig. S6A}). Further, the ultrathin Ga$_2$O$_3$ shell of the droplet is brought in contact with the monolayer area by squeezing the droplet with another piece of a Si/SiO$_2$ wafer on top of the sample with the aid of tweezers ({\color{blue}Fig. S6B}). Careful separation of the small substrate from the sample is crucial for the Ga$_2$O$_3$ sheet to fully delaminate from the gallium and to be transferred on top of monolayer WS$_2$ ({\color{blue}Fig. S6C}). After the transfer, a triangle-shaped CVD-grown monolayer WS$_2$ covered with 2D Ga$_2$O$_3$ can be easily found on a large area of the sample surface. The excess gallium residues can be removed by gently washing the sample in an ethanol or a methanol bath. As result of this process, we achieved direct synthesis of 2D Ga$_2$O$_3$ on top of a large scale area of CVD-grown monolayer WS$_2$ ({\color{blue} Fig. S6D-E}).

\paragraph*{S7: Deterministic transfer method.}
\begin{figure}[H]
\centering
 \includegraphics[width=\linewidth]{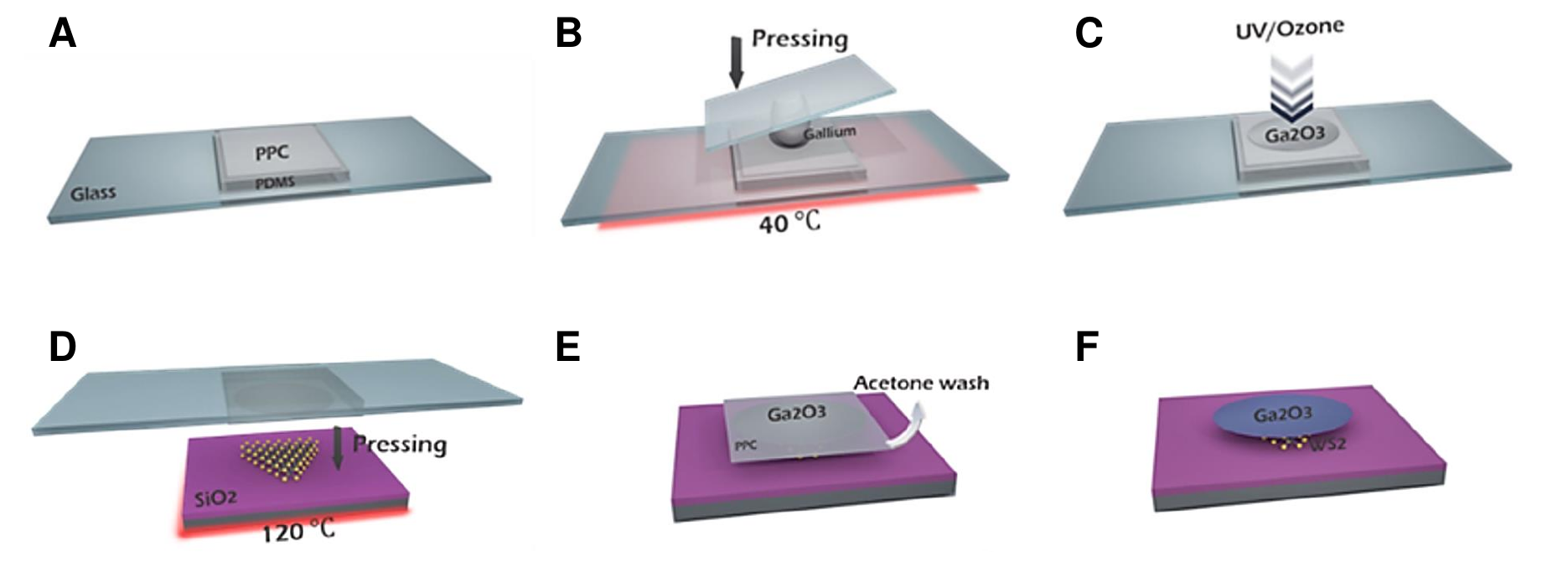}
 \caption*{{\bf Fig. S7:} Schematic illustration of the poly-propylene carbonate (PPC) assisted deterministic transfer of a selected area of 2D Ga$_2$O$_3$ on top of monolayer TMDC: (A) spin-coating of PPC on PMDS and baking at $100~ \mathrm{^{\circ}C}$, (B) synthesis of 2D Ga$_2$O$_3$ sheet on PPC at $40~ \mathrm{^{\circ}C}$, (C) UV/ozone treatment on the Ga$_2$O$_3$ sheet, (D) deterministic transfer of the Ga$_2$O$_3$ sheet on top of monolayer TMDC with a heated transfer stage at $120~ \mathrm{^{\circ}C}$, (E) removal of the PPC residues with acetone, (F) schematic illustration of the transferred 2D Ga$_2$O$_3$ on top of monolayer TMDC.}
 \label{fig:S7}
\end{figure}
Unlike the direct deposition method discussed above (see section S6), the deterministic transfer of liquid metal synthesized 2D Ga$_2$O$_3$ requires an additional sacrificial polymer layer. Therefore, we used poly-propylene carbonate (PPC) with a low thermal stability and a glass transition at around $40~ \mathrm{^{\circ}C}$ \cite{RN566}. First, the PPC in anhydrous anisole solution is spin-coated at $5000~ \mathrm{rpm}$ for $60~ \mathrm{s}$ on top of an elastomer stamp (poly dimethyl siloxane, PDMS) and further baked at $100~ \mathrm{^{\circ}C}$ for $5\ \mathrm{min}$. A glass slide is used to stabilize the PDMS stamp during this process ({\color{blue}Fig. S7A}). Next, 2D Ga$_2$O$_3$ is directly synthesized on top of the PPC layer as described in section S6, but with a synthesis temperature of $40~ \mathrm{^{\circ}C}$ to prevent damage of the temperature sensitive PPC layer ({\color{blue}Fig. S7B}). Using an optical microscop, a clean, gallium-free area can be marked and selected for capping the TMDC monolayer. A UV/ozone treatment is subsequentially applied on the Ga$_2$O$_3$ surface for $15 ~ \mathrm{min}$ to remove the organic residues and to fully oxidize it ({\color{blue}Fig. S7C}). 

Monolayer WS$_2$ is prepared by exfoliation from a commercial grade bulk WS$_2$ crystal (sourced from HQ Graphene) using the conventional scotch tape technique \cite{RN459}, and then transferred on top of a Si/SiO$_2$ substrate. Our exfoliated monolayers are limited to the $\mathrm{\mu m}$ scale due to the limitations of the scotch tape method. Next, the monolayer is aligned with the targeted Ga$_2$O$_3$ area by using a transfer stage with integrated heating ({\color{blue}Fig. S7D}). Before bringing the sample surface in contact with the 2D Ga$_2$O$_3$, the sample is heated up to $60~ \mathrm{^{\circ}C}$, i.e. above the glass phase transition of the PPC. This temperature allows the PPC to follow the sample surface morphology upon contact. After contact, the sample is further heated up to $120~ \mathrm{^{\circ}C}$ to release the PPC \cite{RN454}, with a temperature rate of $10~ \mathrm{^{\circ}C/min}$ to avoid any strain between the substrate, the monolayer, and the Ga$_2$O$_3$. This temperature is held for $60~ \mathrm{s}$ to fully soften the PPC. After that, the PPC layer easily separates from the stamp by slowly lifting it up with the transfer stage and by doing so the transfer of 2D Ga$_2$O$_3$ is achieved ({\color{blue}Fig. S7E}). Subsequently, the sample is placed in a cold acetone bath to dissolve the PPC residues, followed by an isopropyl and an ethanol bath to remove any organic residues left ({\color{blue}Fig. S7F}). Fig. 1E shows that this cleaning procedure does not affect the optical properties of the WS$_2$/Ga$_2$O$_3$ heterostructure.

\subsection*{Photoluminenscence studies on the heterostructures.}
\begin{figure}[H]
\centering
 \includegraphics[width=\linewidth]{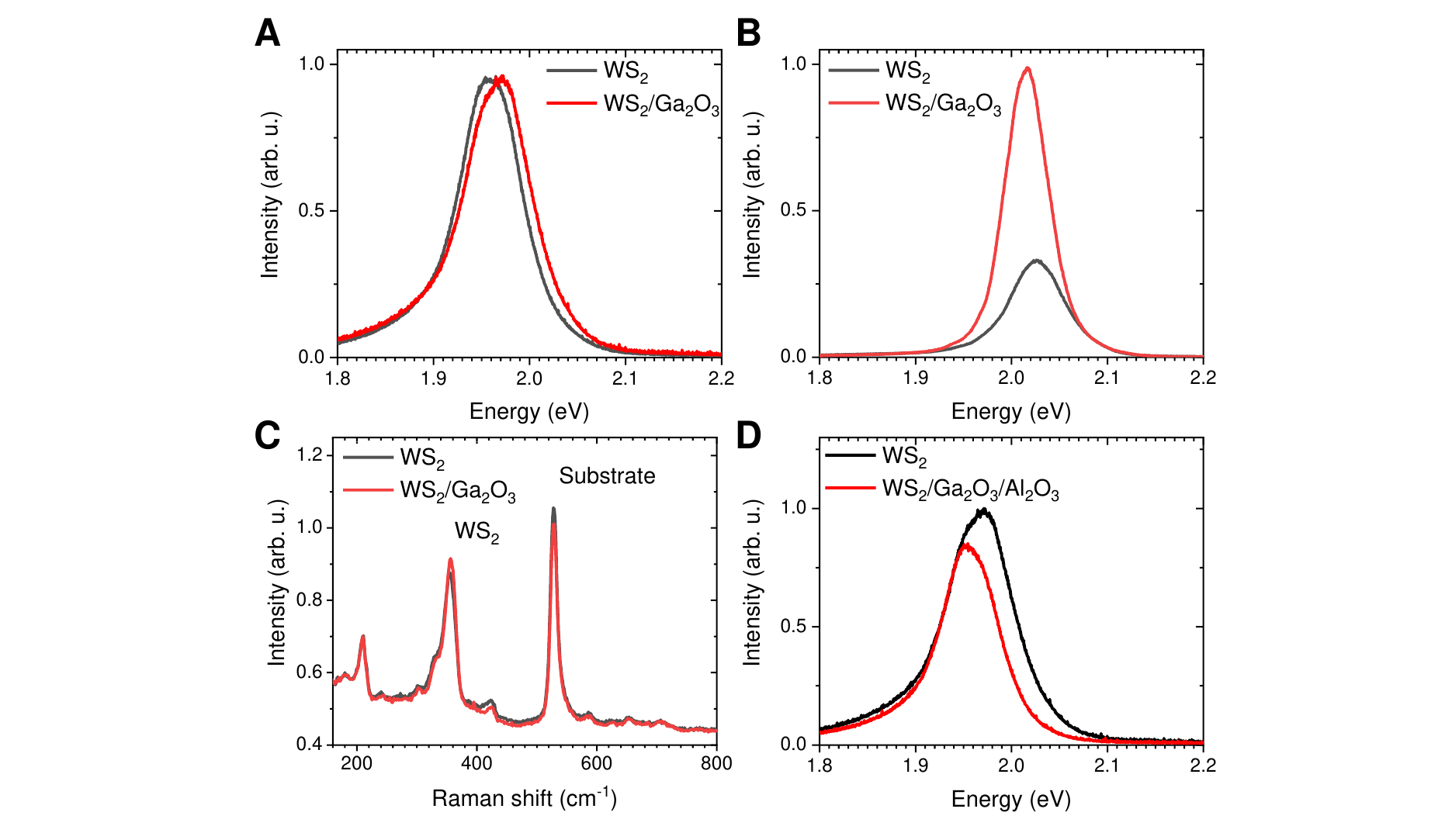}
 \caption*{{\bf Fig. S8:} PL and Raman spectroscopy of CVD-grown monolayer WS$_2$ before and after direct synthesis of 2D Ga$_2$O$_3$: (A) PL spectra at $T\approx 293~\mathrm{K}$, (B) PL spectra at $T\approx 4.3~\mathrm{K}$ and (C)  Raman spectra at $T\approx 4.3~\mathrm{K}$ of uncapped and Ga$_2$O$_3$-capped monolayer WS$_2$; (D) PL spectra of WS$_2$ and WS$_2$/Ga$_2$O$_3$/Al$_2$O$_3$ at $T\approx 293~\mathrm{K}$.}
 \label{fig:S8}
 \end{figure}
\paragraph*{S8: Spectroscopy on CVD-grown WS$_2$ capped by Ga$_2$O$_3$:}
To test the direct synthesis approach (see section S6), we performed the photoluminescence (PL ) and Raman spectroscopy on CVD-grown monolayer WS$_2$ passivated and protected by 2D Ga$_2$O$_3$. {\color{blue}Fig. S8A} and {\color{blue}S8B} show the PL spectra at $T \approx 293~\mathrm{K}$ and at $T\approx4.3~\mathrm{K}$ respectively. In agreement with the results from the main text, the Ga$_2$O$_3$ capping does not quench the monolayer PL at room temperature and causes significant PL intensity enhancement at cryogenic temperatures. Hence, these properties are independent on the transfer technique. The Raman spectra at $T\approx4.3~\mathrm{K}$ ({\color{blue}Fig. S8C}) show that the capping does not affect the optical phonons in the heterostructure, and therefore indicate that the 2D Ga$_2$O$_3$ does not affect the crystal structure of monolayer WS$_2$. 

The PL spectra after EBE deposition of Al$_2$O$_3$ ({\color{blue}Fig. S8D}) further confirm that 2D Ga$_2$O$_3$ capping protects monolayer WS$_2$ against dielectric material deposition, which is the first step towards integrating the CVD-grown monolayers into complex multilayer heterostructures on large scales. For this experiment, we used a relatively low electron-beam energy for the EBE deposition.
 
\paragraph*{S9: Comparison between Ga$_2$O$_3$ and hBN capping of monolayer WS$_2$.}
\begin{figure}[H]
\centering
 \includegraphics[width=\linewidth]{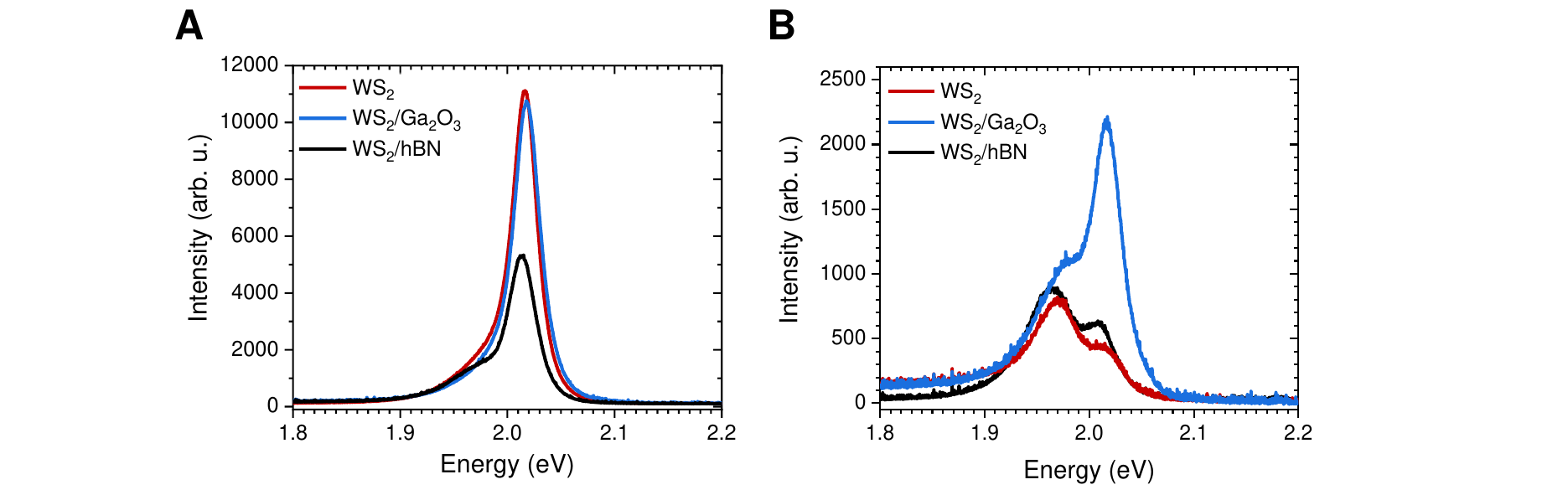}
 \caption*{{\bf Fig. S9:} Comparison of PL spectra of exfoliated monolayer WS$_2$ capped with either 2D Ga$_2$O$_3$ or exfoliated hBN of similar thickness: (A) under ambient conditions, and (B) in vacuum.}
 \label{fig:S9}
\end{figure}
In order to compare the passivating and protecting properties of 2D Ga$_2$O$_3$ and hBN, we prepared a WS$_2$/hBN heterostructure with a thin exfoliated hBN flake. Remarkably, the PL spectrum of the hBN-capped monolayer shows significant quenching of the exciton PL under ambient coniditions, while the Ga$_2$O$_3$ capping fully conserves the monolayer PL ({\color{blue}Fig. S9A}). As explained in the main text, sulphide vacancies in monolayer WS$_2$ can have a large impact on the doping level, the exciton formation, and consequently on the exciton, trion and bound exciton PL \cite{RN504, RN505,RN553}. In air, these vacancies can be passivated with O$_2$ \cite{RN557} or dipolar H$_2$O molecules, and their effect on the PL is reduced. In vacuum, on the other hand, the sulphide vacancy-induced doping strongly affects the PL spectra of monolayer WS$_2$, in which the trion ($E_T \approx 1.96 ~ \mathrm{eV}$) has a pronounced PL feature compared to the exciton ({\color{blue}Fig. S9B}). With the hBN capping, the passivating molecules can be squeezed out of the WS$_2$/hBN interface and the sulphide vacancy-induced doping effect becomes more significant ({\color{blue}Fig. S9A}). In vacuum, the remaining passivating molecules are removed from the interface, which leads to strong sulphide vacancy-induced doping effect  ({\color{blue}Fig. S9B}). On the contrary, with Ga$_2$O$_3$ capping, the sulphide vacancies can be passivated with oxides from the 2D Ga$_2$O$_3$ as explained in the main text, and the emerging donor-type oxide vacancies in the Ga$_2$O$_3$ can be passivated with O$_2$ or the bipolar H$_2$O molecules from air. This passivation of the sulphide vacancies leads to strongly reduced doping effect in vacuum ({\color{blue}Fig. S9B}).

These results demonstrate that 2D Ga$_2$O$_3$ outperforms exfoliated hBN flakes for passivation of commercial grade monolayer WS$_2$ in terms of both scalability and optical performance.

\subsection*{WS$_2$/Ga$_2$O$_3$ at cryogenic temperatures.}
\paragraph*{S10: Intensity-dependent PL of WS$_2$ and WS$_2$/Ga$_2$O$_3$ at cryogenic temperature.}
\begin{figure}[H]
\centering
 \includegraphics[width=\linewidth]{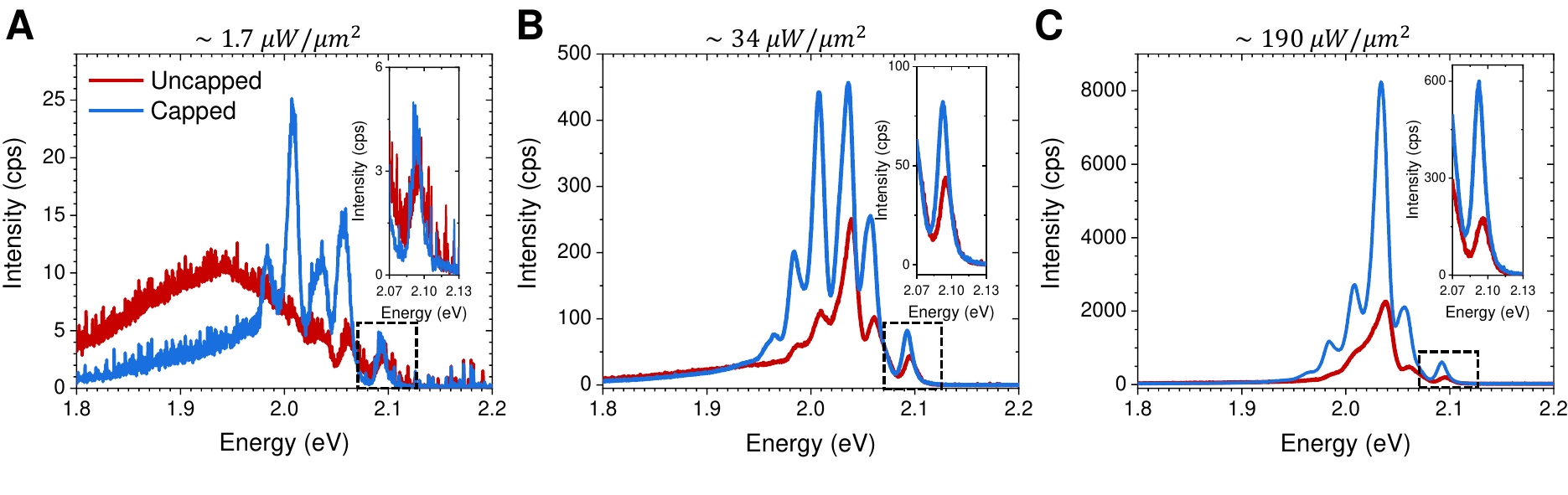}
 \caption*{{\bf Fig. S10:} Low temperature ($T\approx4.3 ~ \mathrm{K}$) PL spectra of the Ga$_2$O$_3$-capped and uncapped monolayer WS$_2$ at three different laser intensities: (A) $I_L \approx 1.7 ~ \mathrm{\mu W/\mu m^2}$, (B) $I_L \approx 34 ~\mathrm{\mu W/\mu m^2}$, and (C) $I_L \approx 190 ~ \mathrm{\mu W/\mu m^2}$, with the magnified exciton PL spectra as insets.}
 \label{fig:S10}
\end{figure}
We performed intensity-dependent PL measurements to investigate the impact of the Ga$_2$O$_3$ capping on the WS$_2$ excitons at cryogenic temperatures ($T\approx4.3 ~ \mathrm{K}$). {\color{blue} Fig. S10A-C} show the PL spectra of both capped and of uncapped monolayer WS$_2$ at three values of laser intensities ($I_L$) spanning more than two orders of magnitude. At $I_L \approx 1.7~ \mathrm{\mu W/\mu m^2}$ the uncapped monolayer WS$_2$ shows strong PL from the bound excitons at $\sim 1.94~\mathrm{eV}$ \cite{RN577}, while the capped monolayer shows strongly enhanced PL from the the many-body exciton complexes at the energies below $\sim 2.05 ~ \mathrm{eV}$. The suppression of the bound exciton formation can be attributed to passivation of the sulphide vacancies with oxides from Ga$_2$O$_3$ (see main text). At $I_L \approx 34 ~ \mathrm{\mu W/\mu m^2}$, the strong enhancement of the many-body complexes is clearly visible in the PL spectrum ({\color{blue} Fig. S10B}), while the PL peak from the biexciton ($E_{XX} \approx 2.04 \ \mathrm{eV}$) experiences a nonlinear increase in intensity compared to the other peaks. Following this nonlinear increase, the biexciton PL dominates the spectrum at high laser intensities $I_L \approx 190 \ \mathrm{\mu W/\mu m^2}$ ({\color{blue}Fig. S10C}), which is in agreement with previous reports \cite{RN502}.

Most importantly, the exciton PL intensities (insets {\color{blue}Fig. S10A-C}) clearly show the laser intensity dependent enhancement in the capped sample, which can be associated with suppression of bound exciton formation, suppressed Auger recombination, and increasing electron-hole injection from the Ga$_2$O$_3$ sheet (see section S12 for details).

\paragraph*{S11: Low temperature cathodoluminescence of WS$_2$/Ga$_2$O$_3$ heterostructure}
\begin{figure}[H]
\centering
 \includegraphics[width=\linewidth]{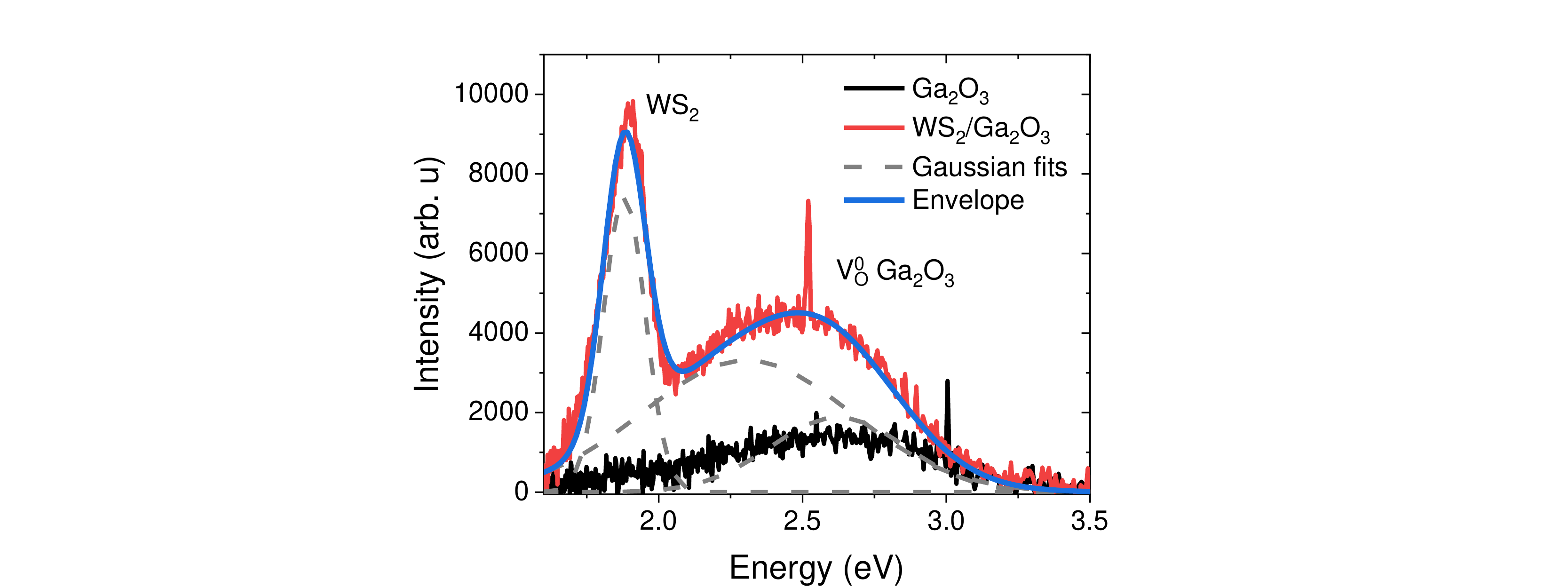}
 \caption*{{\bf Fig. S11:} CL spectra of WS$_2$/Ga$_2$O$_3$ heterostructure and 2D Ga$_2$O$_3$ sheet at $T = \mathrm{70\ K}$, carried out with the electron beam energy of $E\approx 1\ \mathrm{keV}$.}
 \label{fig:S10}
\end{figure}

By using an SEM equipped with a spectrometer, we carried out cathodoluminescence (CL) measurements at $T = 70 \ \mathrm{K}$  to investigate the defect energies in the WS$_2$/Ga$_2$O$_3$ heterostructure. {\color{blue}Fig. S11} shows the CL spectra of both the 2D Ga$_2$O$_3$ sheet and the heterostructure. The spectra were recorded after exciting the samples with an electron beam with the energy of $E\approx 1\ \mathrm{keV}$. The high energy electron beam can excite electrons from the valence band into the vacuum regime or into the conduction band, dependent on the inelastic scattering between the electron beam and the structure. After excitation, the electrons from the conduction band or the electrons from donor type vacancies can recombine with the excited holes in the valence band. If the transitions are optically allowed, this recombination process creates photons with energies corresponding to the energy differences between the electrons and the holes, and a spectrometer can be used to measure the photon energies. The CL spectrum of the WS$_2$/Ga$_2$O$_3$ has two peaks with $E\approx 1.9~ \mathrm{eV}$ and $E \approx 2.5 ~ \mathrm{eV}$ while the CL spectrum of the Ga$_2$O$_3$ has a weaker pronounced feature at $E \approx 2.7 ~ \mathrm{eV}$ ({\color{blue}Fig. S11}). The spectrum of the heterostructure can be well reproduced with three Gaussian peaks with the central energies of $E\approx 1.9~ \mathrm{eV}$, $E \approx 2.25 ~ \mathrm{eV}$ and $E \approx 2.7 ~ \mathrm{eV}$. According to previous reports \cite{RN558,RN507,RN509}, the $E\approx 1.9~ \mathrm{eV}$ peak corresponds to the CL spectrum of monolayer WS$_2$ and the higher energy peaks correspond to oxide vacancies in the Ga$_2$O$_3$. While, according to the Gaussian fits, the CL spectra of the Ga$_2$O$_3$ sheet and the WS$_2$/Ga$_2$O$_3$ heterostructure have a similar intensity at around $E \approx 2.7 ~ \mathrm{eV}$, the emission from the $E \approx 2.25 ~ \mathrm{eV}$ vacancies is strongly enhanced in the heterostructure ({\color{blue}Fig. S11}). This is an evidence of a significantly higher density of the new donor-type oxide vacancies in the heterostructure compared to the bare Ga$_2$O$_3$ sheet.

Together with the suppression of bound exciton formation (see {\color{blue}Fig. S10A} in section S10), these results suggests that the oxides from the Ga$_2$O$_3$ sheet fill the sulphide vacancies in WS$_2$, which leads to the increase of the donor-type oxide vacancy density in the heterostructure, while the doping level is not affected (see {\color{blue}Fig. 2B} in the main text).  
 
\paragraph*{S12: Theoretical model for the exciton generation in the WS$_2$/Ga$_2$O$_3$ heterostructure.}
\begin{figure}[H]
\centering
 \includegraphics[width=\linewidth]{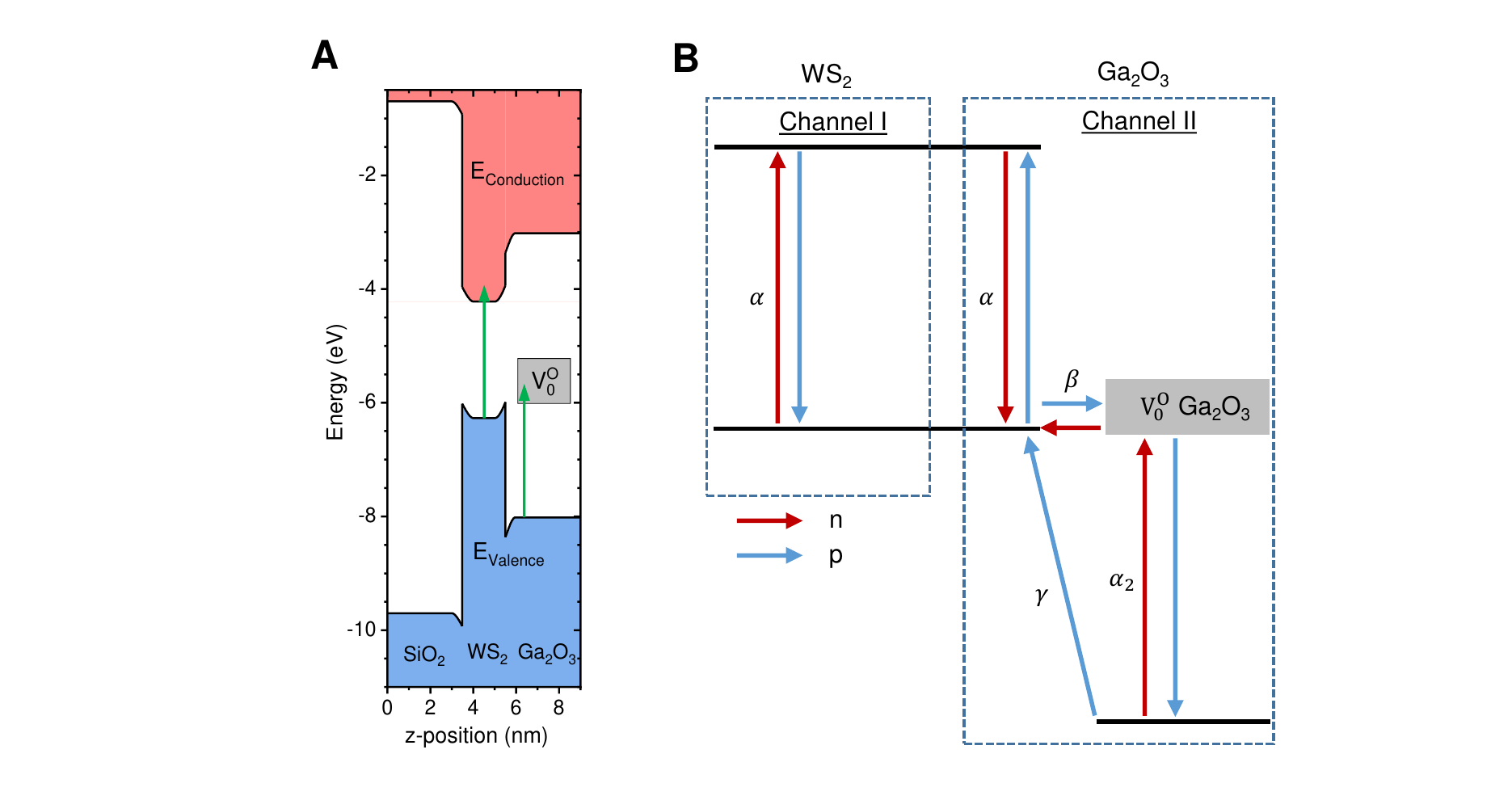}
 \caption*{{\bf Fig. S12.1:} Direct and indirect exciton generation in the WS$_2$/Ga$_2$O$_3$ heterostructure. (A) Band diagram of the SiO$_2$/WS$_2$/Ga$_2$O$_3$ heterostructure including the optically active oxide vacancies from section S11. (B) Schematic band diagram of the WS$_2$/Ga$_2$O$_3$ heterostructure with transitions for direct electron-hole pair generation in monolayer WS$_2$ (Channel I) and indirect electron-hole pair generation via the oxide vacancies in the Ga$_2$O$_3$ sheet (Channel II). $n$ and $p$ are the excited electron and hole densities respectively, $\alpha$ is the generation rate of excited electrons and holes in monolayer WS$_2$, $\alpha_2$ is the generation rate of Ga$_2$O$_3$ electrons into the oxide vacancies, $\beta$ is the tunnelling rate from the vacancy electrons into the WS$_2$ valence band, and $\gamma$ is the thermalization and tunnelling rate for the excited Ga$_2$O$_3$ holes.}
 \label{fig:S12}
\end{figure}
To model the intensity dependent exciton PL intensities for both the capped and uncapped monolayer WS$_2$, we constructed a rate equation model, which accounts for the direct exciton-hole generation in monolayer WS$_2$, as well as for the indirect electron-hole generation via the oxide vacancies in the Ga$_2$O$_3$ sheet. {\color{blue}Fig. S12.1A} shows the band diagram of the SiO$_2$/Ga$_2$O$_3$ heterostructure including the optically active oxide vacancies with the energies extracted from the CL measurement ({\color{blue}Fig. S11}). According to this band diagram, the energies of the donor type oxide vacancies are above the WS$_2$ valence band maximum. Hence, the donor electrons can tunnel into the excited WS$_2$ holes which is crucial for the process described here. 

After tunnelling, a laser source with sufficient energy can replenish the empty oxide vacancies by exciting the Ga$_2$O$_3$ valence band electrons, and further excite the tunnelled electrons from the Ga$_2$O$_3$ into the WS$_2$ conduction band. Finally, the excited electrons can couple to the tunnelled and thermalized holes from the Ga$_2$O$_3$ valence band forming excitons. The efficiency of this two-step process scales with the  number of excited WS$_2$ holes. Therefore, with growing intensity, $I_L$, this effect leads to an increasing power law exponent $k$ for the exciton intensity $I_X \propto I_L^k$. In other words, the excited WS$_2$ holes effectively catalyse the electron-hole generation in Ga$_2$O$_3$ which competes against the direct exciton formation. Hence, this process delays the exciton formation, and ultimately the electron-hole recombination, leading to a significant increase of the exciton density at high excitation intensities. Moreover, since the Ga$_2$O$_3$ is highly transparent to the laser used in our experiment with $E_P\approx 2.33 \ \mathrm{eV}$, this process has a negligible effect on the direct exciton generation in monolayer WS$_2$. 

{\color{blue}Fig. S12.1B} schematically shows the band diagram of the WS$_2$/Ga$_2$O$_3$ heterostructure with the highlighted transitions for the direct electron-hole pair generation in the monolayer WS$_2$ (Channel I) and for the indirect electron-hole generation via the oxide vacancies (Channel II). For the rate equations, we only consider the formation of exctions, and disregard the formation of trions and other many-body complexes. Therefore, we only take into account the contribution of the donor electrons to the exciton formation and not to the formation of charged particles, e.g. trions.

The process depicted in Fig. {\color{blue}S12.1B} can be described by the following rate equations: 
\begin{align}
\frac{dn}{dt}&=\alpha I_L +\beta N_0 p \alpha I_L - c_X np, \\
\frac{dp}{dt}&=\alpha I_L + \gamma p_{Ga_2O_3} + \beta N_0 p \alpha I_L - \beta N_0 p - c_Xnp, \\
\frac{dN_0}{dt}&=\alpha_2 I_L - \beta N_0 p, \\
\frac{dp_{Ga_2O_3}}{dt}&= \alpha_2 I_L - \gamma p_{Ga_2O_3},
\end{align}
with the excitation intensities $I_L$,  the electron-hole excitation rate $\alpha$ in the monolayer WS$_2$, the density of excited electrons  $n$ and holes $p$ in the monolayer WS$_2$, the donor electron excitation rate $\alpha_2$, the density of excited donor electrons $N_0 \in [0, N]$ ($N\widehat{=}$   total donor density), the number of excited holes $p_{Ga_2O_3}$ in the Ga$_2$O$_3$ sheet, the donor electron tunnelling rate $\beta$ into the excited WS$_2$ holes, the tunnelling and thermalization rate $\gamma$ for $p_{Ga_2O_3}$ into the monolayer WS$_2$, and the exciton formation rate $c_X$.

In order to account for exciton annihilation processes (e.g., Auger recombination) in this model, we recalculate the exciton density by using an effective loss factor:
\begin{equation}
\bar{X} = X(1 - \Gamma(X)),
 \end{equation}
where $X = c_X np$. Here $\Gamma(X)\in(0,1)$ phenomenologically describes the exciton density dependent two-particle annihilation processes \cite{RN473} between multiple excitons \cite{RN555,RN556,RN559}, between excitons and defects  \cite{RN555,RN556,RN560}, and between free carriers and defects \cite{RN555}, assuming an equilibrium between generated and annihilated excitons at high exciton densities $X>N_x$ above the saturation density $N_X$ what leads to $\lim_{P \to \infty} \bar{X} = N_X$: 
\begin{equation}
\Gamma(X) = \frac{X}{X + N_X}.
\end{equation}
The factor $\Gamma(X)$ accounts only for two particle annihilation, but not for other excitation intensity dependent processes, e.g. exciton Mott transition \cite{RN565}, biexciton formation \cite{RN502} and heating effects, which would lead to additional effective quantum yield losses at high excitation intensities. To verify this phenomenological model, we fitted $\bar{X}$ to the experimental intensity dependent PL measurements, and achieved a good fit to both the exciton intensities $I_X$ and to the extracted power law exponent $k$ with $N_X = 3,900 ~\mathrm{arb. u.}$ for the uncapped sample (see {\color{blue}Fig. 2C}). Hence, the phenomenological loss factor $\Gamma(X)$ adequately describes the effective annihilation processes we observe in the uncapped WS$_2$ monolayer within our experimental intensity range. In the experiment, $\mathrm{arb. u.}$ corresponds to $\mathrm{counts*pixel^{-1}*s^{-1}}$ in our spectrometer camera. 

Assuming that all the generated excitons $c_X np = c_X n^2$ contribute to the exciton PL, we simulated the excitation intensity dependent exciton PL for the rate equations including the $\Gamma(X)$ function (eq. (6)), without ($N = 0$) and with ($N > 0$) the active Channel II in {\color{blue}Fig. S12.2B}.

The results for excitation intensities spanning six orders of magnitudes are presented in {\color{blue}Fig. S12.2A-F}. For the simulations, we used the fitted parameters for the exciton PL intensities $I_X(I_L)$ of the WS$_2$/Ga$_2$O$_3$ heterostructure from {\color{blue}Fig. 2C}, but for simplicity treated them as dimensionless: $\alpha = 10$, $\alpha_2 = 4\times10^{-4} \alpha$, $c_X = 0.2 \alpha$, $N = 10^3$, and $\beta = 1\times10^{-4}$. Here, the zero-offset of $I_X(I_L)$ is described by $\alpha$, and the line shape of $I_X(I_L)$ by the ratio $\alpha_2/ \alpha$. The value of this ratio is in agreement with the high optical transmissivity of the Ga$_2$O$_3$ sheet compared to monolayer WS$_2$, and we achieved a good fit to the experimental data with this value. The other free parameters dictate whether the vacancies $N_0$ are exhausted at high excitation intensities for the indirect exciton generation process ($N_0 = 0$). This is unlikely within our experimental excitation intensity range, since the trion to exciton ratio is still significant at high intensities in the WS$_2$/Ga$_2$O$_3$ heterostructure ({\color{blue}Fig. S10C}), which justifies the selected values for $c_X$, $N$ and $\beta$. For comparison, we calculated the exciton intensities for $N_X = \infty$, $N_X= 3,900$ and $N_X = 110,000$. {\color{blue}Fig. S12.2A} and {\color{blue}S12.2D} show the corresponding excitation intensity dependent $\Gamma(X)$ functions with deactivated and activated Channel II respectively, showing that $\Gamma(X)$ asymptotically approaches 1 at high excitation intensities $I_L$ for a finite $N_X$, as observed phenomenologically. 

First, we discuss the simulation results without the exciton annihilation processes ($N_X = \infty$). With the deactivated Channel II, the exciton intensity $I_X$ strictly follows a power law $I_X \propto I_L^k$ with $k=1$ (see  {\color{blue}Fig.S12.2B-C}), confirming that Channel I of our model describes the direct photon-exciton transition leading to the exciton generation in monolayer WS$_2$. With Channel II active, the power law exponent $k$ transitions from $k=1$ to $k=2$ with increasing excitation intensity (see {\color{blue}Fig. S12.2E-F}), which is in agreement with our experimental results (see {\color{blue}Fig. 2C}).

With finite $N_X$ and without Channel II active, one can clearly see the deviation from the strict power law at high excitation intensities ({\color{blue}Fig. S12.2C} and {\color{blue}S12.2E}) which is in agreement with previous reports on exciton annihilation processes in monolayer TMDCs \cite{RN473} and with our measurements. With smaller $N_X = 3,900$, the power law is more affected, as expected. 

 \begin{figure}[H]
\centering
 \includegraphics[width=\linewidth]{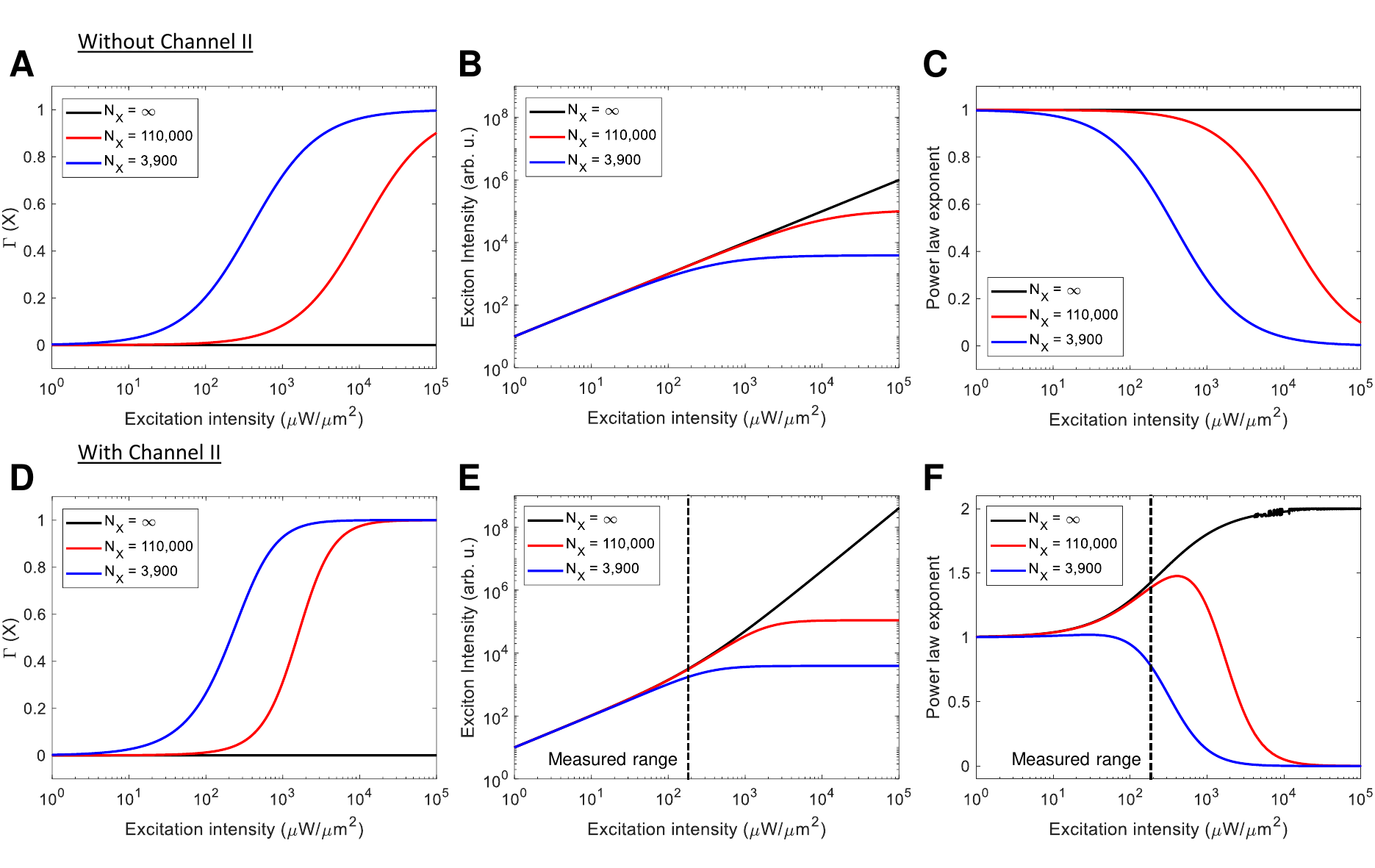}
 \caption*{{\bf Fig. S12.2:} Simulated excitation intensity dependent WS$_2$ exciton PL intensities for the unified rate equation model from Fig. S12.1. including the loss factor $\Gamma(X)$ (eq. (6)), with $N_X = \infty$ (black), $N_X = 110,000$ (red) and $N_X= 3,900$  (blue), for the direct and indirect exciton generation in the WS$_2$/Ga$_2$O$_3$ heterostructure (A-C) without  and (D-F) with channel II: (A), (D) excitation intensity-dependent $\Gamma(X)$ function, (B,E) excitation intensity dependent exciton intensities, (C,F) corresponding power law exponent $k$. The marked area in (E) and (F) corresponds to the experimental intensity range presented in Fig. 2C.}
 \label{fig:S12.2}
\end{figure}

With Channel II active, the exciton annihilation processes strongly limits the nonlinear growth of the exciton intensity and the maximum exciton density is reached at lower excitation intensities, $I_L$, compared to the case with deactivated Channel II. With $N_X = 3,900$, which is the fitted maximum exciton density for the uncapped monolayer ({\color{blue}Fig. 2C} in the main text), the power law exponent $k$ drops below 1 at  excitation intensities above $\sim 50 ~ \mathrm{\mu W/\mu m^2}$ (see {\color{blue}Fig. 12.2F}), showing that the maximum exciton density of the capped sample has to be higher than that of the uncapped monolayer. With an increased saturation density of $N_X = 110,000$, the power law exponent peaks around $k \approx 1.5$ and drops strongly at  $I_L > 1,000 ~\mathrm{\mu W/\mu^2 m}$, at intensities more than one order of magnitude above the experimental intensity range (marked area in {\color{blue}Fig. S12.2E-F}). This means that due to the Ga$_2$O$_3$-induced nonlinear exciton generation in monolayer WS$_2$, the maximum exciton density is reached at lower excitation intensities compared to the uncapped monolayer. The competition of the hole recycling process with the exciton formation processes allows for significantly higher exciton densities at lower excitation intensities. 

By using the model with both Channel I and Channel II, and by modifying the exciton density $X$ with the phenomenological loss factor $\Gamma(X)$, we achieved a good fit for the measured exciton intensities ({\color{blue}Fig. 2C} in the main text), with $N_X = 3,900$ for the uncapped monolayer and with $N_X = 110,00$ for the capped monolayer. This indicates that the saturation density in the WS$_2$/Ga$_2$O$_3$ heterostructure is enhanced by $\sim 28$ times compared to the uncapped sample.

%Materials and Methods\\
%Supplementary Text\\
%Figs. S1 to S3\\
%Tables S1 to S4\\
%References \textit{(4-10)}

% For your review copy (i.e., the file you initially send in for
% evaluation), you can use the {figure} environment and the
% \includegraphics command to stream your figures into the text, placing
% all figures at the end.  For the final, revised manuscript for
% acceptance and production, however, PostScript or other graphics
% should not be streamed into your compliled file.  Instead, set
% captions as simple paragraphs (with a \noindent tag), setting them
% off from the rest of the text with a \clearpage as shown  below, and
% submit figures as separate files according to the Art Department's
% instructions.

%\noindent {\bf Fig. 1.} Please do not use figure environments to set
%up your figures in the final (post-peer-review) draft, do not include graphics in your
%source code, and do not cite figures in the text using \LaTeX\
%\verb+\ref+ commands.  Instead, simply refer to the figure numbers in
%the text per {\it Science\/} style, and include the list of captions at
%the end of the document, coded as ordinary paragraphs as shown in the
%\texttt{scifile.tex} template file.  Your actual figure files should
%be submitted separately.

\bibliographystyle{Science}
%\thebibliography{Manuscript}

\end{document}